\documentclass[12pt]{article}

\usepackage{graphicx}
\usepackage{float,color}
\usepackage{amsmath}
\usepackage{amssymb}
\usepackage{cite}               
\newcommand{\dd}{\mathrm{d}}
\newcommand{\expo}{\mathrm{e}}

\begin{document}

\begin{center}
\Large 
From short-range repulsion to Hele-Shaw problem in a model of tumor growth
\end{center}
\begin{center}
Sebastien Motsch\footnote{Sebastien Motsch: Physical Sciences Building, A-Wing Rm. 836, School of Mathematical and Statistical Sciences, Arizona State, University Tempe, AZ 85287-1804, smotsch@asu.edu
},  Diane Peurichard\footnote{Diane Peurichard : Faculty of Mathematics, Wien University, Oskar-Morgenstern Platz, 1090 Wien, {diane.peurichard@univie.ac.at
}}
\end{center}
\begin{abstract}

  We investigate the large time behavior of an agent based model describing tumor growth. The {\it microscopic} model combines short-range repulsion and cell division. As the number of cells increases exponentially in time, the microscopic model is challenging in terms of computational time. To overcome this problem, we aim at deriving the associated {\it macroscopic} dynamics leading here to a porous media type equation. As we are interested in the long time behavior of the dynamics, the macroscopic equation obtained through usual derivation method fails at providing the correct qualitative behavior (e.g. stationary states differ from the microscopic dynamics). We propose a modified version of the macroscopic equation introducing a density threshold for the repulsion. We numerically validate the new formulation by comparing the solutions of the {\it micro-} and {\it macro-} dynamics. Moreover, we study the asymptotic behavior of the dynamics as the repulsion between cells becomes singular (leading to non-overlapping constraints in the microscopic model). We manage to show formally that such an asymptotic limit leads to a Hele-Shaw type problem for the macroscopic dynamics.

The macroscopic model derived in this paper therefore enables to overcome the problem of large computational time raised by the microscopic model and stays closely linked to the microscopic dynamics.
\end{abstract}

\thanks{This work has been supported by the NSF grants DMS-1515592 and RNMS11-07444 (KI-Net). The first author wishes to thank A. Mellet and Velasquez for helpful discussions.}



\section{Introduction}
\setcounter{equation}{0}

One of the main difficulties in the modeling of complex systems such as fish schooling or tumor growth is the lack of fundamental laws. It is unknown how two {\it agents} (e.g. cells, pedestrians, birds) interact, we only have access to the result of the interactions. But there is always one rule that agents have to satisfy: two agents cannot {\it overlap}, i.e. they cannot occupy the same position in space. Despite the simplicity of this rule, non-overlapping constraints have several intriguing effects and raise several challenges both analytically and numerically, as they induce non-convex problems. Several methods have been proposed to encompass non-overlapping constraints. At the {\it microscopic} level (i.e. agent-based models), a common method is to introduce (short-range) repulsion dynamics: two cells move away from each other when they are too close. At the {\it macroscopic} level (i.e. Partial Differential Equations (PDE)), non-overlapping can be expressed as a density constraint, i.e. the density has to stay below a given threshold. The goal of this work is to link the two descriptions starting from a simple model of tumor growth.

In the literature, the problem of non-overlapping is ubiquitous in the modeling of collective behavior. When pedestrians are crossing \cite{helbing_social_1985,moussaid_traffic_2012,maury_discrete_2011} or when birds flock together \cite{ballerini_empirical_2008}, avoidance of neighbors is always necessary. Usually, this rule is modeled by a repulsion interaction \cite{aoki_simulation_1982,reynolds_flocks_1987,couzin_collective_2002,fetecau_swarm_2011}. At the macroscopic level, the density constraint is a key factor and is responsible for instance in the formation of car traffic jam \cite{berthelin_model_2008,berthelin_traffic-flow_2008,berthelin_model_2012}. The effect of congestion leads to numerous challenging mathematical models such as non-linear diffusion \cite{burger_nonlinear_2010,burger_stationary_2014} and two-phase flows \cite{bouchut_hierarchy_2000,degond_self-organized_2013}. More generally density constraints have been studied for fluid models in \cite{berthelin_existence_2002,degond_numerical_2011,labbe_free_2013,bresch_singular_2014,perrin_free/congested_2015,degond_congestion_2010,degond_macroscopic_2014}. Incompressibility constraints have also been analyzed using optimal transport theory \cite{maury_macroscopic_2010}. Finally, the derivation of macroscopic equations from microscopic dynamics have been extensively studied in the case of repulsion interaction \cite{burger_aggregation_2007,morale_interacting_2005,oelschlager_large_1990,kipnis_hydrodynamics_1989} and in case of volume exclusion \cite{bruna_excluded-volume_2012,bruna_diffusion_2014}.

In cancer modeling specifically, space occupancy is of critical importance. The dynamics of tumor growth can either be described through agent based models \cite{byrne_individual-based_2009,leroy_leretre_etude_2014,degond_macroscopic_2014}, fluid dynamics \cite{roose_mathematical_2007,lowengrub_nonlinear_2010,bresch_computational_2010} or simply reaction-diffusion equations \cite{harpold_evolution_2007,swanson_virtual_2003}. However, all of these models share a common feature: density effects are encoded in the cell behavior (e.g. cell division or necrosis depending on the density). Of main importance for this paper is the derivation of Hele-Shaw type problem for tumor growth \cite{perthame_heleshaw_2014,perthame_derivation_2014,perthame_traveling_2014,mellet_hele-shaw_2015}

The focus of this paper is to explore the effects of density constraint in a simple model of tumor growth. The starting point is an agent-based model (referred to as the {\it microscopic} dynamics) which combines short-range repulsion and cell division. Different behaviors are observed depending on key parameters (i.e. strength repulsion, growth rate). We are interested in the large time behavior of the model.  Since agent-based models are not easily tractable (dealing with millions of cells), we investigate how to capture these characteristic behaviors through a {\it macroscopic} description (i.e. using PDE).

In a first attempt, we derive a macroscopic model using the weak equation satisfied by the so-called empirical distribution. We observe that micro- and macro- dynamics have drastically different behaviors: the microscopic solutions converge to a stationary state, whereas the macroscopic solutions keep spreading in space. Here, as we are interested in the long time behavior, there is no guarantee that both micro- and macro- dynamics remain close, even if the number of particles $N$ becomes large. Moreover, for short range repulsion, Dirac masses are not stable \cite{balague_nonlocal_2013} which also explains why the macroscopic dynamics diverges from the particle simulations. The difficulty comes from the 'local' range of interaction: repulsion should only apply when particles are 'close enough'. But the notion of {\it closeness} is lost when the dynamics is described through a density distribution $\rho$. For this reason, we modify the macroscopic dynamics to {\it de-activate} repulsion at low-density. This modification allows to retrieve solutions with the same qualitative behavior as the microscopic dynamics.

The second contribution of the paper is to explore an asymptotic limit when the repulsion between cells becomes 'singular'. Formally, such an asymptotic limit leads to non-overlapping constraints in the microscopic dynamics. At the macroscopic level, we show (formally) that this asymptotic limit leads to a Hele-Shaw type problem. Several numerical tests comparing the micro- and macro- dynamics are performed in this limit. In particular, explicit solutions of the Hele-Shaw problem are found and are in excellent agreement with the microscopic dynamics.

Since this manuscript is a first attempt to link an agent based model for tumor growth with a Hele-Shaw type problem, several questions remain unanswered and require further investigations. For instance, from a theoretical viewpoint, the rigorous derivation of the Hele-Shaw problem is still an open problem. This will require precise estimates depending on the size and number of cells. From a modeling viewpoint, one should consider more elaborate dynamics, such as adding nutriments and/or a more complex rules for cell division. The challenge will be then to derive the corresponding macroscopic limit. Numerically, solving the microscopic dynamics with non-overlapping constraint is computationally demanding. It is difficult to find efficient algorithms to solve the non-overlapping constraint with a large number of cells. One could finally compare the microscopic dynamics with the Hele-Shaw problem in a more complex environment (e.g. presence of necrosis, vascularity etc).

The manuscript is organized as follows. In section 2, we present the microscopic dynamics (i.e. agent-based model combining short-range repulsion and cell division) and identify numerically three distinctive behaviors. In section 3, we introduce a macroscopic dynamics (PDE) associated with the agent-based model and include a density constraint to match the solutions of the microscopic dynamics. In section 4, we investigate analytically and numerically the limit as the repulsion between cells becomes singular, leading to a Hele-Shaw type problem. We draw conclusions and future works in section 5.

\section{Agent-based model}

\subsection{Short-range repulsion and cell division}

We consider a dynamical system of $N$ cells moving in a plane with short range repulsion. Each cell is modeled as a 2D sphere of center ${\bf x}_i\in\mathbb{R}^2$ and radius $R>0$. Cells repulse each other at short distance according to the following dynamics:
\begin{equation}
  \label{eq:micro_kpp2}
  \dot{{\bf x}}_i = -\sum_{j=1,j\neq i}^N \phi_{ij}({\bf x}_j-{\bf x}_i) \qquad \text{with} \quad \phi_{ij} = \phi\left(\left|\frac{{\bf x}_j-{\bf x}_i}{2R}\right|^2\right),
\end{equation}
where $\phi \geq 0$ is the interaction function. As we intend to model short-range repulsion, the support of $\phi$ is $[0,1]$ which implies that cells do not interact if their centers are at distance greater than $2R$. We consider the following cell-cell repulsion function:
\begin{equation}
  \label{eq:phi_example}
  \phi(s) = \left\{
    \begin{array}{ll}
      \frac{1}{s} - 1 & \text{, if }  0<s\leq1\\
      0 & \text{, otherwise}.
    \end{array}
  \right.
\end{equation}
Notice that $\phi$ is singular at the origin.

{\bf remark}
  There is an energy associated with this dynamics, namely:
  \begin{equation}
    \label{eq:energy_micro}
    \mathcal{E}(\{{\bf x}_i\}_i) = -\sum_{i<j} \Phi\left(\left|\frac{{\bf x}_j-{\bf x}_i}{2R}\right|^2\right),
  \end{equation}
  with $\Phi$ an anti-derivative of $\phi$. The functional $\mathcal{E}$ is decaying along the solutions $\{{\bf x}_i(t)\}_i$ of \eqref{eq:micro_kpp2}. More precisely, using that $\nabla_{{\bf x}_i}\mathcal{E} = -\frac{1}{4R^2} \dot{{\bf x}}_i$, we find:
  \begin{displaymath}
    \frac{\dd}{\dd t} \mathcal{E}(\{{\bf x}_i(t)\}) = \sum_i\nabla_{{\bf x}_i}\mathcal{E}\cdot\dot{{\bf x}}_i = - \frac{1}{4R^2} \sum_i |\dot{{\bf x}}_i|^2 \;\;\leq\;\; 0.
  \end{displaymath}
  This property is used to build an adapted numerical scheme (see appendix \ref{sec:appendix_A}), the time step $\Delta t$ is chosen such that the energy is always decaying numerically.

In addition to the short-range repulsion, the dynamics is coupled with {\it cell division}. Each cell divides at a given frequency $\mu$ and creates a new cell in its neighborhood. Mathematically, this birth process is modeled as a Poisson process: over a time period $\Delta t$, a cell divides with probability $1-\expo^{-\mu\Delta t}$. Once a cell $k$ divides, a new cell $k^*$ is created at the position:
\begin{equation}
  \label{eq:cell_division}
  {\bf x}_{k^*} = {\bf x}_k + R\cdot\varepsilon,
\end{equation}
where $\varepsilon$ is a random variable uniformly distributed on the unit disc. The small displacement $R\cdot\varepsilon$ has been introduced to avoid the singularity of the interaction function $\phi$ at the origin (i.e. ${\bf x}_{k^*}-{\bf x}_k=0$ otherwise).

\begin{table}
  \centering
  \begin{tabular}{l|c|r}
    \hline
    Initial number of cells    &  $N$    &  $100-500$   \\
    Cell-cell repulsion    &  $\phi$    &  Eq. \eqref{eq:phi_example} \\
    Range repulsion        &  $R$    &  $2\cdot10^{-1}$ \\
    Growth rate            &  $\mu$    &  $5\cdot 10^{-2}$ \\
    \hline
  \end{tabular}
  \caption{Parameters used for the Microscopic model \eqref{eq:micro_kpp2}.}
  \label{tab:param_ABM}
\end{table}

\subsection{Qualitative behaviors}

To illustrate the agent-based model, we explore the dynamics in three different scenarios. These simulations will be used in the next chapter as {\it test cases} for the validation of the derived macroscopic models.

First, we explore the dynamics without cell division (i.e. $\mu=0$) and investigate its long-time behavior. The system converges to a stationary distribution where cells organize into a lattice configuration where no cells overlap (i.e. $|{\bf x}_i-{\bf x}_j|\geq2R$). In the second setting, cell division is turned on and the dynamics no longer converges to a stationary state. Finally, we keep cell division but enforce that cells do not overlap at each discrete time step. This corresponds formally to an asymptotic limit where the repulsion force between cells becomes infinite.

\subsubsection{Repulsion only}

\begin{figure}[H]
  \centering
  \includegraphics[scale=.85]{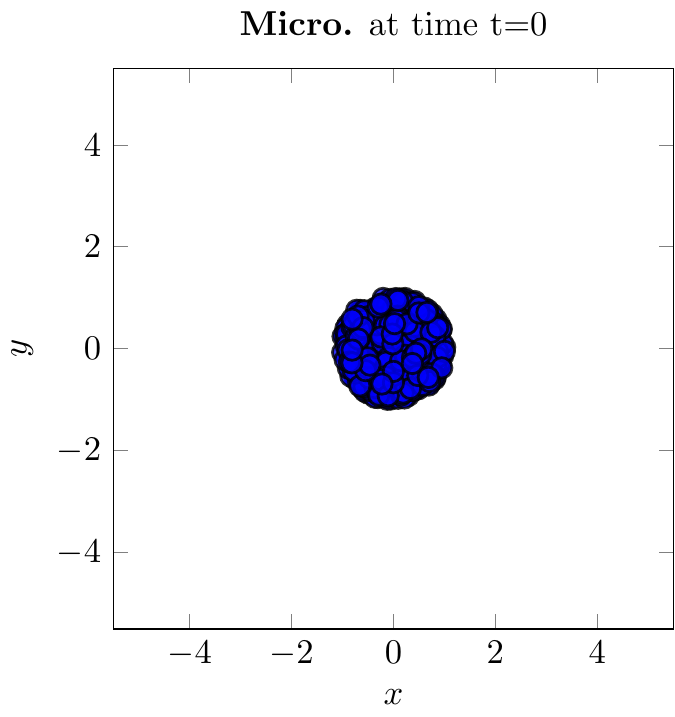}
  \includegraphics[scale=.85]{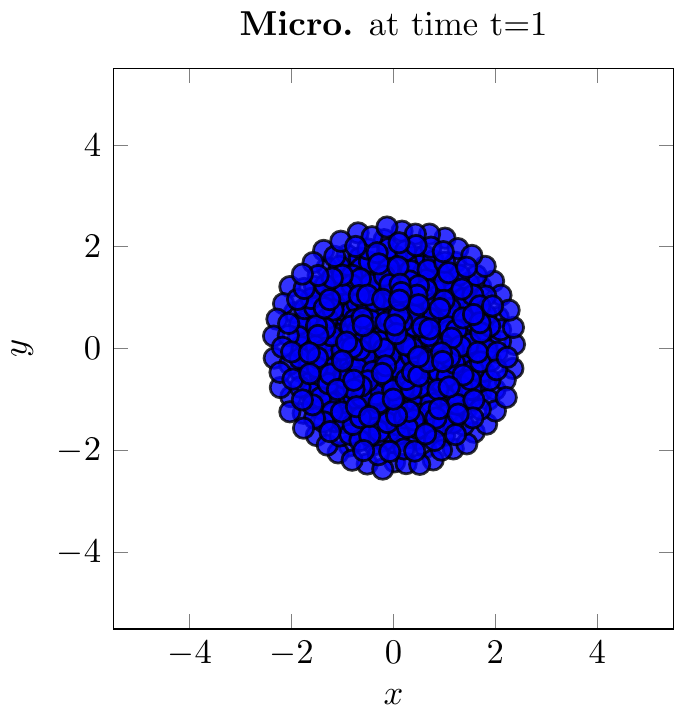} \\
  \includegraphics[scale=.85]{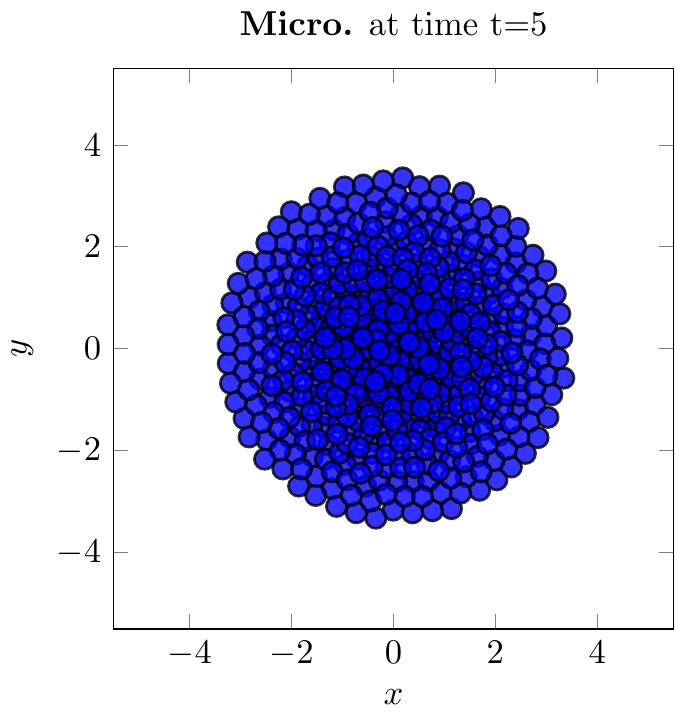}
  \includegraphics[scale=.85]{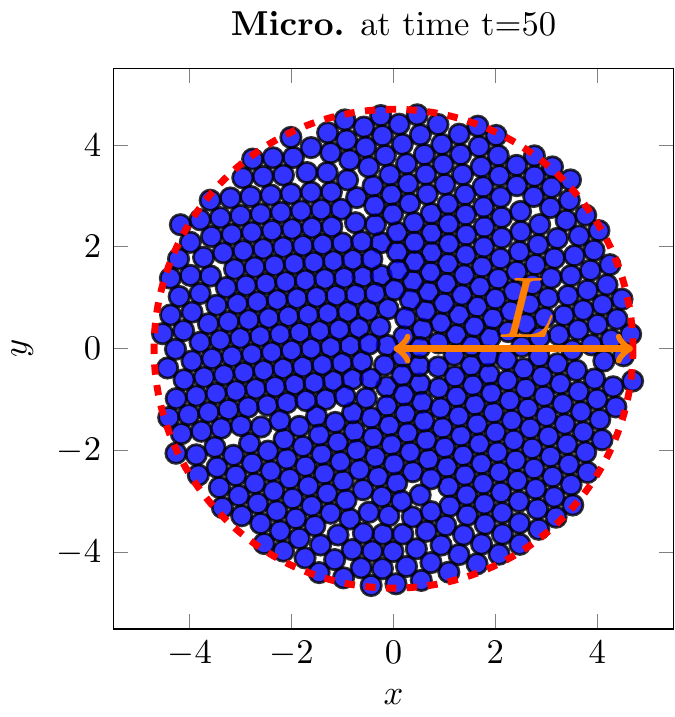}
  \caption{The dynamics \eqref{eq:micro_kpp2} with repulsion only. Initially (top-left), cells are distributed in the unit circle. The cells spread over time ($t=1,5$ time units) until reaching an equilibrium at $t=50$ time units (bottom-right). We draw in red the diameter circle predicted by the maximum packing number of circles.  Parameters: $N=500$ cells, radius $R=.2$, $\Delta t=10^{-1}$.}
  \label{fig:Micro_repulsion}
\end{figure}

Without cell division, cells keep pushing each other until all the cells are at a distance greater than $2R$ from each other. Starting from an initial configuration with $N=500$ cells distributed uniformly on the unit disc (see Fig.~\ref{fig:Micro_repulsion} top-left), the cell population spreads in space until it reaches a stationary state given by an 'hexagonal lattice' structure. Initially, cells disperse fastly due to the large cell overlapping rate, and the dispersion slows down as cells separate from each others. After $t=50$ time units (Fig.~\ref{fig:Micro_repulsion} bottom-right), the system is close to a stationary state. It is straight-forward to see that in $\mathbb{R}^2$ the set of stationary states for the dynamics \eqref{eq:micro_kpp2} is given by:
\begin{displaymath}
  \mathcal{C} = \big\{ \{{\bf x}_i\}_i\,: \;\; |{\bf x}_i-{\bf x}_j|\geq2R \big\}.
\end{displaymath}
Thus, the dynamics \eqref{eq:micro_kpp2} can be seen as a penalizing method to enforce the non-overlapping constraints $|{\bf x}_i-{\bf x}_j|\geq2R$ for all $i,j$. As the set $\mathcal{C}$ is non-convex, enforcing such constraint is challenging. There exist efficient algorithms to enforce that the dynamics would never exit the constraint domain $\mathcal{C}$ \cite{maury_time-stepping_2006,maury_discrete_2011}, i.e. cells will not overlap at any time. However in the model of interest in this paper, cell-cell overlapping cannot be prohibited at any time because of the cell division process (which generates cell-cell overlapping regularly in time). 

We can estimate the radius $L$ of the disc surrounding the stationary state. Suppose that the cells are in a configuration of optimal arrangement. In $\mathbb{R}^2$, the highest density among all the possible packing arrangements for circles of constant radius is $\pi/2\sqrt{3}$. This result, due to the works of Gauss \cite{gaus_s_besprechung_1831} corresponds to an hexagonal lattice. In our setting, this leads to:
\begin{displaymath}
  L = \sqrt{N\cdot\frac{2\sqrt{3}}{\pi}}\cdot R \approx 4.70
\end{displaymath}
since we use $N=500$ cells with radius $R=.2$ space units. Fig. \ref{fig:Micro_repulsion}-bottom right shows an excellent agreement between the circle of radius $L$ (centered in the cell population's center of mass) and the boundary of the cell population. Note however that some cells are slightly beyond the circle line due to the fact that cells are not enforced to be in optimal packing arrangement.

\begin{figure}[H]
  \centering
  \includegraphics[scale=.65]{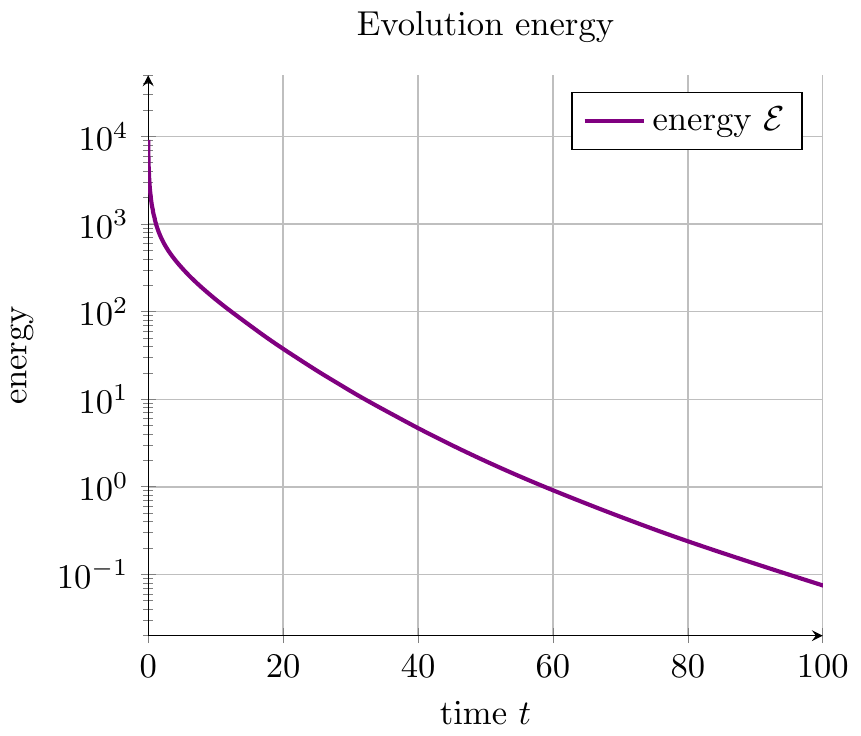} \quad
  \includegraphics[scale=.65]{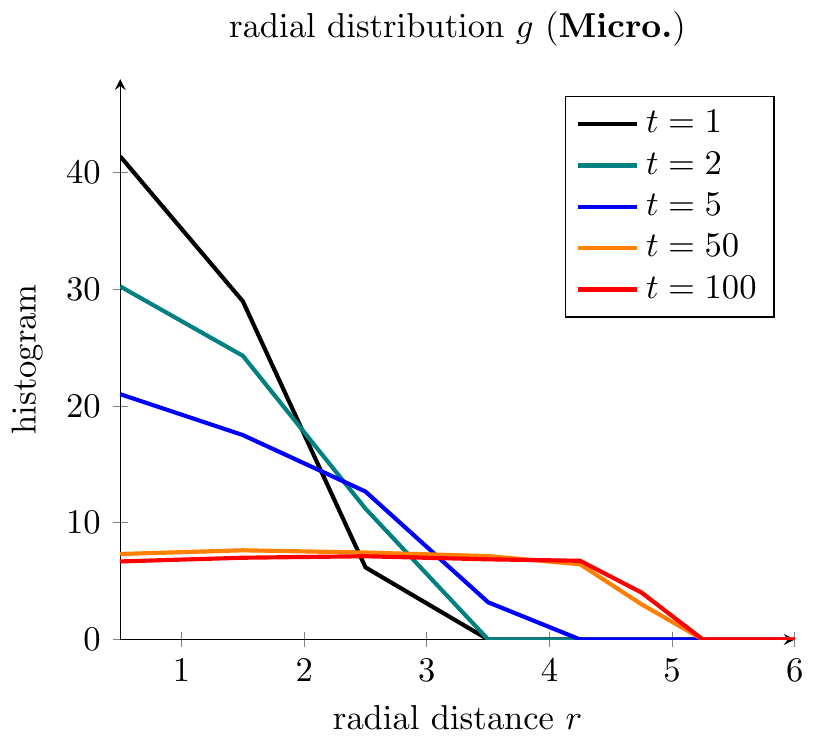}
  \caption{{\bf Left:} Energy $\mathcal{E}$ \eqref{eq:energy_micro} for the simulation of Fig. \ref{fig:Micro_repulsion}. The decay is exponential in time. {\bf Right:} Radial distribution $g(r)$ \eqref{eq:radial} for the same simulation. As the stationary state is given by a uniform distribution of cell on a disc, the distribution $g$ converges to a plateau.}
  \label{fig:Micro_repulsion_e_rhoR}
\end{figure}

To measure the convergence to an equilibrium state, we can compute the energy \eqref{eq:energy_micro} over time. In Fig. \ref{fig:Micro_repulsion_e_rhoR}-left, we observe that the energy is decaying exponentially in time. At $t=50$, the energy has already decayed by $3$ orders of magnitude.

Since the cells are spreading radially, an efficient way to characterize the dynamics is to compute the radial distribution $g(r)$, which gives the average number of cells on a disc of size $r$:
\begin{equation}
  \label{eq:radial}
  g(r) \Delta r = \frac{1}{2\pi r} \#\{r-\Delta r/2\leq|{\bf x}|\leq r+\Delta r/2\}.
\end{equation}
In Fig. \ref{fig:Micro_repulsion_e_rhoR}-right, we observe that the radial distribution $g(r)$ converges to a plateau distribution (except near its boundary $r \approx 4.7$ unit). 

\subsubsection{Repulsion and cell division}

\begin{figure}[H]
  \centering
  \includegraphics[scale=.74]{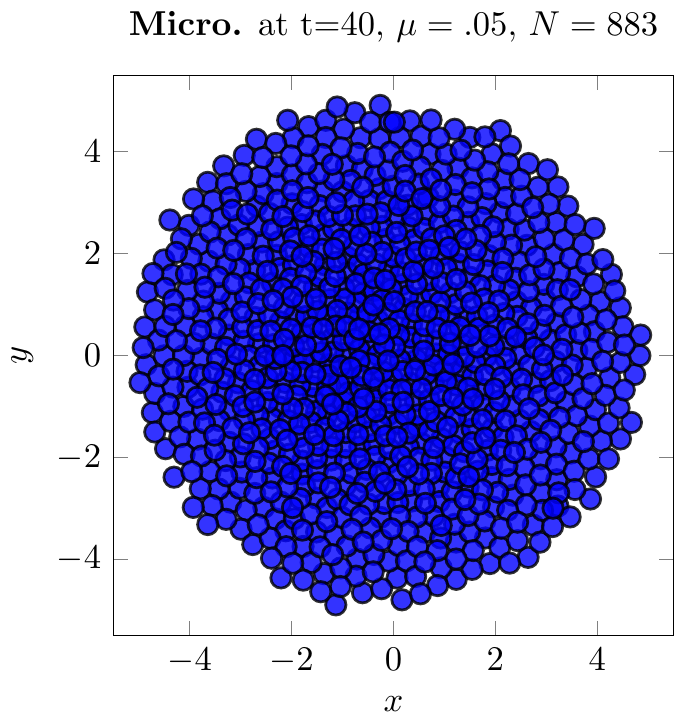} \quad
  \includegraphics[scale=.74]{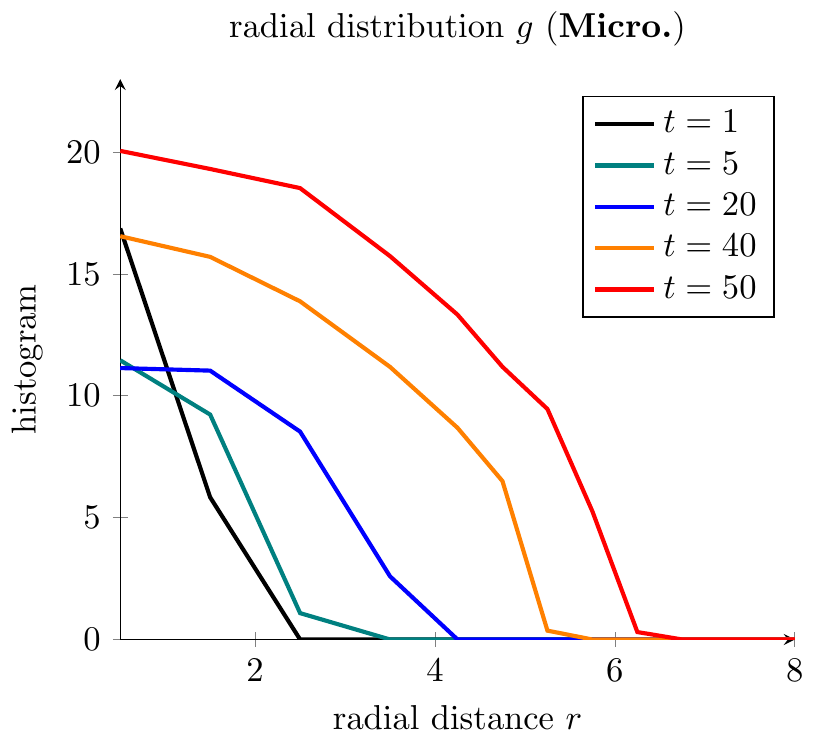}
  \caption{{\bf Left:} Distribution of cells at $t=40$ with cell division activated ($\mu=.05$). Initially, only $100$ cells are distributed on the unit circle. {\bf Right:} Radial distribution $g$ of cells at several times. The distribution is no longer converging to a stationary distribution.}
  \label{fig:Micro_repulsion_growth}
\end{figure}

In this section, we turn on cell division with rate $\mu=.05$. The number of cells is therefore increasing exponentially in time. In Fig. \ref{fig:Micro_repulsion_growth}, we plot a snapshot of a simulation starting with $N=100$ cells distributed on the unit disc. We observe that the center of the cell population is denser than its rim. The radial distribution of cells $g$ shows that the cell density in the center first decays (up to $t\approx20$ unit time) due to cell-cell repulsion and then increases in time. In contrast to the previous simulation (Fig.~\ref{fig:Micro_repulsion}), the dynamics do not converge to a stationary state. The radial distribution keeps growing and expanding. 

These results highlight the competition between cell repulsion -which tends to decrease the cell density- and cell division -which increases the local cell density-, and show that there exists a transition time after which cell repulsion is not strong enough to prevent the cell population to expand in space.  

\subsubsection{Non-overlapping and cell division}
\label{sec:micro_nonoverlapping}

Finally, we would like to implement the (non-convex) constraint that cells should not overlap. Numerically, for each time step $\Delta t$, we let the repulsion dynamics \eqref{eq:micro_kpp2} runs until it reaches (numerically) a stationary state. More precisely, we set up a tolerance $\epsilon>0$ ($\epsilon=.01$ in the simulations) and run the repulsion dynamics \eqref{eq:micro_kpp2} until $|{\bf x}_i^{n+1}-{\bf x}_i^n|\leq\epsilon\cdot R$ for all $i$. Once the stationary state is reached, cell division occurs \eqref{eq:cell_division} and the repulsion dynamics is once again activated until it reaches a stationary state. Following this process, the distribution of cells remains close to a non-overlapping configuration at each observational time. Note that such a setting amounts to consider that the characteristic time of cell repulsion is much smaller than the one of cell division: cells reach the non overlapping configuration between two cell division events.

\begin{figure}[H]
  \centering
  \includegraphics[scale=.7]{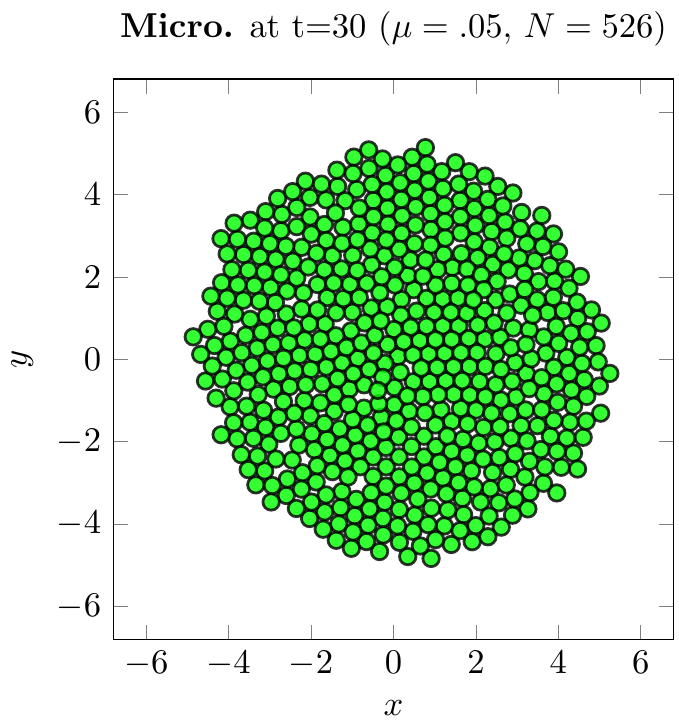} \quad
  \includegraphics[scale=.7]{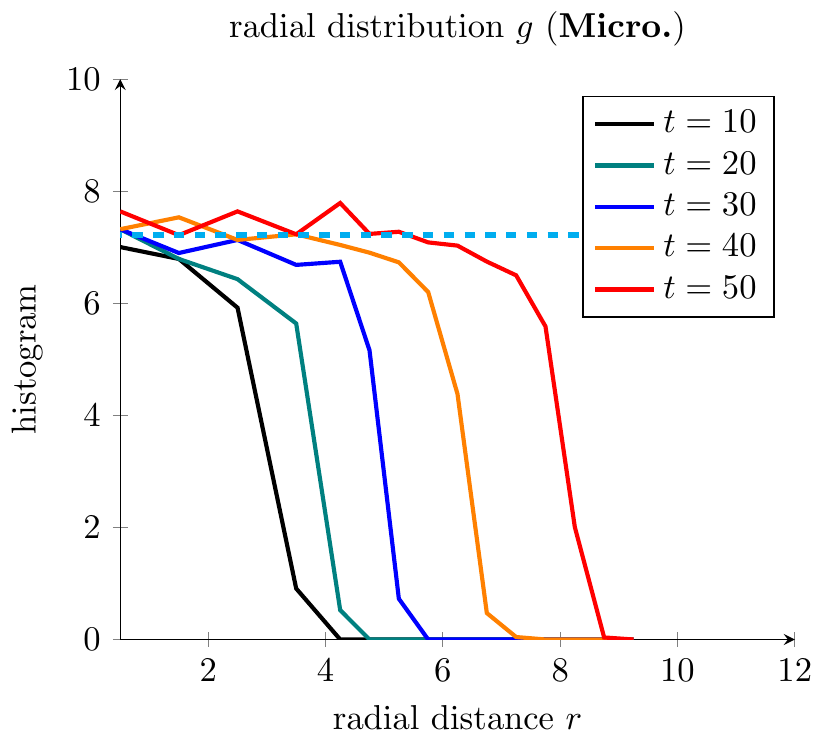}
  \caption{{\bf Left:} Cells at $t=30$ with cell division and non-overlapping constraint activated. {\bf Right:} The radial distribution of cells $g$ is expanding in space but remains close to the maximum packing number (i.e. $\rho_*\approx7.217$) on its support.}
  \label{fig:Micro_repulsion_growth_hulk}
\end{figure}

In Fig. \ref{fig:Micro_repulsion_growth_hulk}, we plot the result of this algorithm starting with the same initial condition as in Fig. \ref{fig:Micro_repulsion_growth}. The comparison between Fig. \ref{fig:Micro_repulsion_growth} and Fig. \ref{fig:Micro_repulsion_growth_hulk} shows that cell diffusion in space is much faster in this regime than in the one of previous section. These are expected results since cell repulsion is much faster when cell non-overlapping is treated as a constraint than when treated as a force.  Moreover, the radial distribution remains close to the maximum packing number ($\rho_* = \frac{\pi}{2\sqrt{3}} \frac{1}{\pi R^2} \approx7.217$). Since the total density is increasing exponentially, we deduce that the front of the distribution is also growing exponentially.


\section{Macroscopic model}

\subsection{Unstable approach}


We would like to analyze the {\it microscopic} dynamics \eqref{eq:micro_kpp2} from a {\it macroscopic} point of view. With this aim, we would like to derive the PDE associated with the repulsion dynamics \eqref{eq:micro_kpp2}. The standard method is to consider the so-called empirical distribution:
\begin{equation}
  \label{eq:empirical}
  \rho({\bf x},t) = \sum_i \delta({\bf x}-{\bf x}_i(t)), 
\end{equation}
where $\{{\bf x}_i(t)\}_i$ is solution of the dynamical system \eqref{eq:micro_kpp2}. To find the equation satisfied by the empirical distribution, we integrate $\rho$ against a test function $\varphi$ and take the time derivative. One deduces that if $\mu=0$ (repulsion only), $\rho$ satisfies (weakly) the following equation:
\begin{displaymath}
  \partial_t \rho + \nabla_{\bf x} \cdot(G[\rho]\rho) = 0,
\end{displaymath}
with
\begin{equation}
  \label{eq:G_rho}
  G[\rho]({\bf x}) = -\int_{{\bf y}\in\mathbb{R}^2} \phi\left(\left|\frac{{\bf x}-{\bf y}}{2R}\right|^2\right)({\bf y}-{\bf x}) \rho({\bf y})\, \dd {\bf y}.
\end{equation}
Combining repulsion and cell division leads to the following dynamics:
\begin{equation}
  \label{eq:PDE_integral}
  \partial_t \rho + \nabla_{\bf x} \cdot(G[\rho]\rho) = \mu\rho.
\end{equation}

{\bf remark}
  We do not normalize the empirical distribution \eqref{eq:empirical} by $1/N$ in order to keep the information of the total number of cells. The total mass is essential at the particle level to determine the size of the support of the stationary state. If one would like to study the asymptotic limit $N\to\infty$, one would have to normalize the empirical distribution by $N$ and consider the limit $R\to0$.

\begin{figure}[H]
  \centering
  \includegraphics[scale=.65]{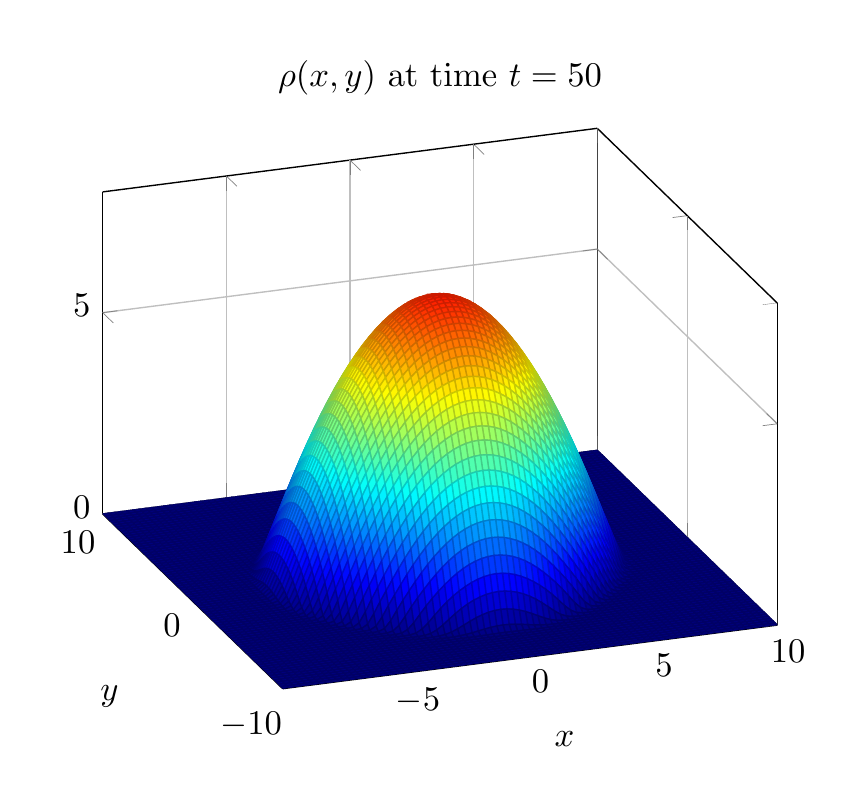}
  \includegraphics[scale=.65]{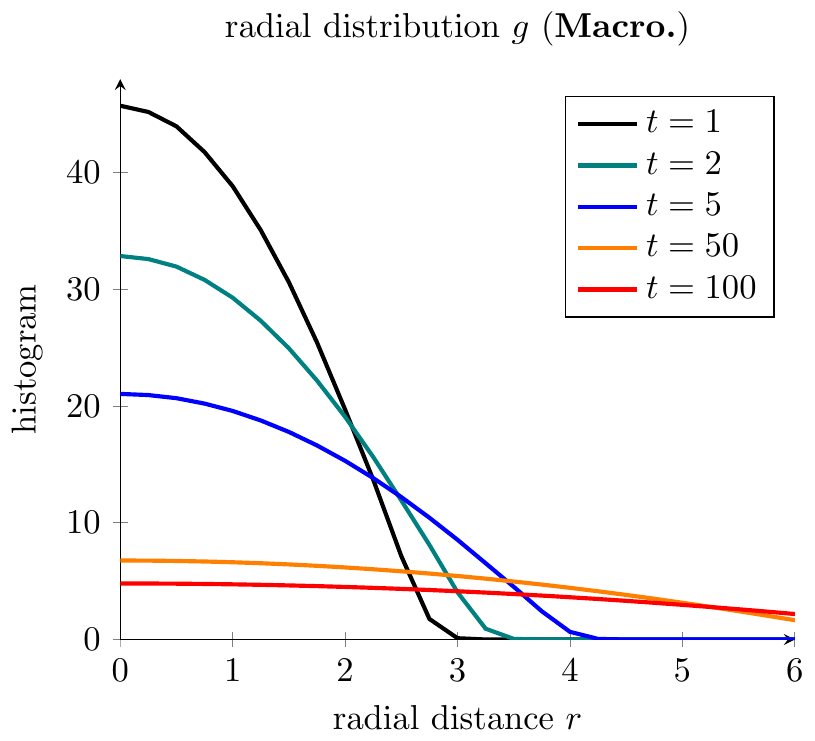}
  \caption{{\bf Left:} The solution $\rho({\bf x},t)$ of the macroscopic dynamics \eqref{eq:G_rho}\eqref{eq:PDE_integral} with initial condition \eqref{eq:rho_0} at $t=50$. {\bf Right:} Radial plot at several time. The density keeps spreading in space. Parameters: $\Delta x=\Delta y=.25,\,\Delta t=5\cdot10^{-2}$.} 
  \label{fig:macro_conv}
\end{figure}

As an illustration of the macroscopic dynamics, we perform a numerical simulation in the same setting as section 2.2.1 (i.e. no cell division $\mu=0$). The initial cell density $\rho_0$ corresponds to $500$ cells distributed on the unit circle:
\begin{equation}
  \label{eq:rho_0}
  \rho_0({\bf x}) = \left\{
    \begin{array}{ll}
      \frac{500}{\pi}  & \text{if } |{\bf x}|\leq 1\\
      0  & \text{if } |{\bf x}|> 1
    \end{array}
  \right.
\end{equation}
We plot in Fig. \ref{fig:macro_conv}-left the density $\rho({\bf x},t)$ at  $t=5$ time units. We observe that the cell density spreads more in the macroscopic model than in the microscopic one (Fig.~\ref{fig:Micro_repulsion}). More specifically, the radial distribution $g$ defined by (using the radial symmetry of the configuration):
\begin{displaymath}
  g(t,r) = \rho(t,x=r,y=0),
\end{displaymath}
does not converge to a stationary state but rather converges to zero in time (even though the total mass is conserved). Note that this is not observed with the microscopic model (Fig.~\ref{fig:Micro_repulsion_e_rhoR}), for which the cell density rather converges to a finite value.

Since the macroscopic dynamics \eqref{eq:G_rho}\eqref{eq:PDE_integral} is derived from the microscopic dynamics \eqref{eq:micro_kpp2}, the discrepancies between the two dynamics require some explanations. There are two factors to take into account. First, the correspondence between the two dynamics (micro- and macro-) can only be proven as the number of cells $N$ tends to infinity \cite{oelschlager_large_1990}. Here, the number of cells is supposed to be fixed and equal to $N=500$. Secondly, we have investigated the large time behavior of the dynamics (i.e. $t\to\infty$), and since there is no 'uniform' bounds in time between the microscopic and the macroscopic dynamics, one cannot guarantee that the two solutions will remain close.

Concerning our specific dynamics of short-range repulsion, there is a key observation: Dirac masses are not stable for the macroscopic dynamics. Starting from $\rho_0({\bf x})$ a perturbation of a Dirac mass in Eq. \eqref{eq:G_rho}\eqref{eq:PDE_integral}, the solutions $\rho({\bf x},t)$ will diffuse in space and thus departs from the Dirac distribution. A more detailed analysis of the stability of 'shell solutions' is provided in \cite{balague_nonlocal_2013}. Therefore, even though the macroscopic dynamics  have as a weak solution the empirical distribution \eqref{eq:empirical}, this solution is {\it unstable}. Thus, it will not be observed numerically. Notice that in the case of an attractive potential (i.e. $\phi<0$ in our settings), a Dirac distribution would be stable.

\subsection{Stabilizing method}

As we have shown previously, the simulations of the macroscopic dynamics \eqref{eq:G_rho}\eqref{eq:PDE_integral} do not match the solutions of the microscopic dynamics \eqref{eq:micro_kpp2}. One explanation is that Dirac masses are not stable solutions for the dynamics \eqref{eq:PDE_integral}. In this section, we would like to modify Eq. \eqref{eq:PDE_integral} to better assess the microscopic dynamics (e.g. similar stationary solutions). 

\begin{figure}[H]
  \centering
  \includegraphics[scale=1.2]{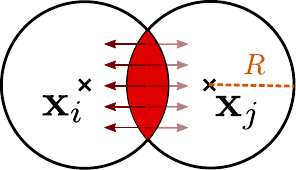}
  \caption{The two cells located at ${\bf x}_i$ and ${\bf x}_j$ interact when they are at a distance less than $2R$ from each other. On this region, the modified empirical distribution $\widetilde{\rho}$ \eqref{eq:modified_empirical} is larger than the threshold $\rho_{*}=\frac{1}{\pi R^2}$.}
  \label{fig:repulsion_sioux}
\end{figure}

The main idea is as follows: cells do not interact once they are at a distance greater than $2R$ from each other. Unfortunately, the information of inter-cell distance (i.e. $|{\bf x}_i-{\bf x}_j|$) is lost when we describe a system with a density distribution $\rho({\bf x})$. To overcome this challenge, lets assume that cells are described by discs of size $R$ rather than point masses. Mathematically, this assumption corresponds to smoothing the empirical distribution with a convolution against an indicator function $\varphi_R= \frac{1}{\pi R^2} \chi_{B(0,R)}$:
\begin{equation}
  \label{eq:modified_empirical}
  \widetilde{\rho}({\bf x},t) = \rho*\varphi_R = \frac{1}{\pi R^2} \sum_{i=1}^N \chi_{B({\bf x}_i(t),R)}({\bf x}).
\end{equation}
The normalization $\pi R^2$ ensures that the total number of cells remains $N$. Cell-cell interactions then occur only in the region where cells overlap each others (see Fig.~\ref{fig:repulsion_sioux}), and this region is described as $\{\widetilde{\rho} > \rho_*\}$ with $\rho_*=\frac{1}{\pi R^2}$. Thus, repulsion should be activated only when the density is larger than the threshold $\rho_*$.


\begin{figure}[H]
  \centering
  \includegraphics[scale=.6]{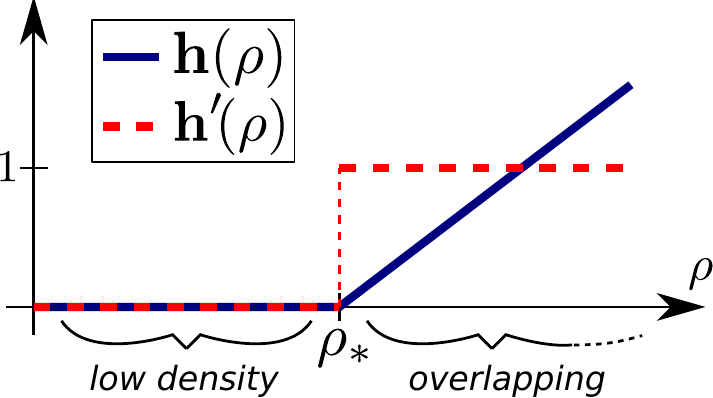}
  \caption{The function ${\bf h}(\rho)=(\rho-\rho_*)^+$ used to modify the dynamics in low density region.}
  \label{fig:example_h}
\end{figure}

For this reason, we  modify the macroscopic dynamics \eqref{eq:G_rho}\eqref{eq:PDE_integral} to encompass that no interaction should occur if the density $\rho$ is below a certain threshold. We fix a threshold $\rho_*$ and consider the following dynamics:
\begin{equation}
  \label{eq:PDE_integral_sioux}
  \partial_t \rho + \nabla_{\bf x} \cdot(\overline{G}[\rho]\rho) = \mu\rho,
\end{equation}
with
\begin{equation}
  \label{eq:G_rho_sioux}
  \overline{G}[\rho]({\bf x}) = -\int_{{\bf y}\in\mathbb{R}^2} \phi\left(\left|\frac{{\bf x}-{\bf y}}{2R}\right|^2\right)({\bf y}-{\bf x}) {\bf h}(\rho({\bf y}))\, \dd {\bf y},
\end{equation}
and ${\bf h}$ is a piecewise linear function (see Fig. \ref{fig:example_h}):
\begin{equation}
  \label{eq:h}
  {\bf h}(\rho) =  \left\{
    \begin{array}{ll}
      0 & \text{ if } \rho<\rho_*,\\
      \rho-\rho_* & \text{ if } \rho \geq \rho_*.
    \end{array}
  \right.
\end{equation}

{\bf remark}
If there is no low density region (i.e. $\rho({\bf x})>\rho_*$ for all ${\bf x}$), then:
\begin{eqnarray*}
  \overline{G}[\rho]({\bf x}) &=& -\int_{y\in\mathbb{R}^2} \phi\left(\left|\frac{{\bf x}-{\bf y}}{2R}\right|^2\right)({\bf y}-{\bf x}) \big(\rho({\bf y}) -\rho_*\big)\, \dd {\bf y} \\
                           &=&  G[\rho]({\bf x}),
\end{eqnarray*}
by symmetry. Therefore, the dynamics \eqref{eq:PDE_integral_sioux}\eqref{eq:G_rho_sioux} only modifies the previous dynamics \eqref{eq:G_rho}\eqref{eq:PDE_integral} inside and nearby the low density region.  


\subsection{Numerical illustrations}

We illustrate numerically the dynamics \eqref{eq:PDE_integral_sioux}\eqref{eq:G_rho_sioux} using the same setting as section 3.1 (no cell division $\mu=0$ and initial condition given by Eq. \eqref{eq:rho_0}). We fix the threshold $\rho_*=\frac{c}{\pi R^2}$ where $c=\pi/2\sqrt{3}\approx.907$ is the 'packing number' of circles in $\mathbb{R}^2$. We observe in Fig. \ref{fig:macroSioux_conv} that the distribution $\rho({\bf x},t)$ stops spreading when the density is lower than the threshold $\rho_*$. As a result, the density converges to a stationary distribution given by a characteristic function on a circle of radius $L\approx4.7$. This result is confirmed by the evolution of the radial distribution $g$ (see Fig.  \ref{fig:macroSioux_conv}-right). The distribution $g$ becomes constant up to a radius of $4.5$ unit space which is consistent with the microscopic dynamics (see Fig.~\ref{fig:Micro_repulsion_e_rhoR}).

Similarly, we compare the micro- and macro- simulations turning on cell division with rate $\mu=.05$. In Fig. \ref{fig:macroSioux_conv}, we plot the radial distribution of the solutions of the macroscopic dynamics (continuous line) and compare them with the microscopic dynamics (dashed line). Without the stabilizing method, the macroscopic solution of \eqref{eq:G_rho}\eqref{eq:PDE_integral} (left) diffuses too fast. However Fig. \ref{fig:macroSioux_conv} right shows the good agreement between the PDE \eqref{eq:PDE_integral_sioux}\eqref{eq:G_rho_sioux} and the microscopic model. There are some discrepancies at the boundary of the cell density support, due to the stochastic aspect of cell division in the microscopic model. One could obtain a better agreement by taking an average over several microscopic simulations.

\begin{figure}[H]
  \centering
  \includegraphics[scale=.65]{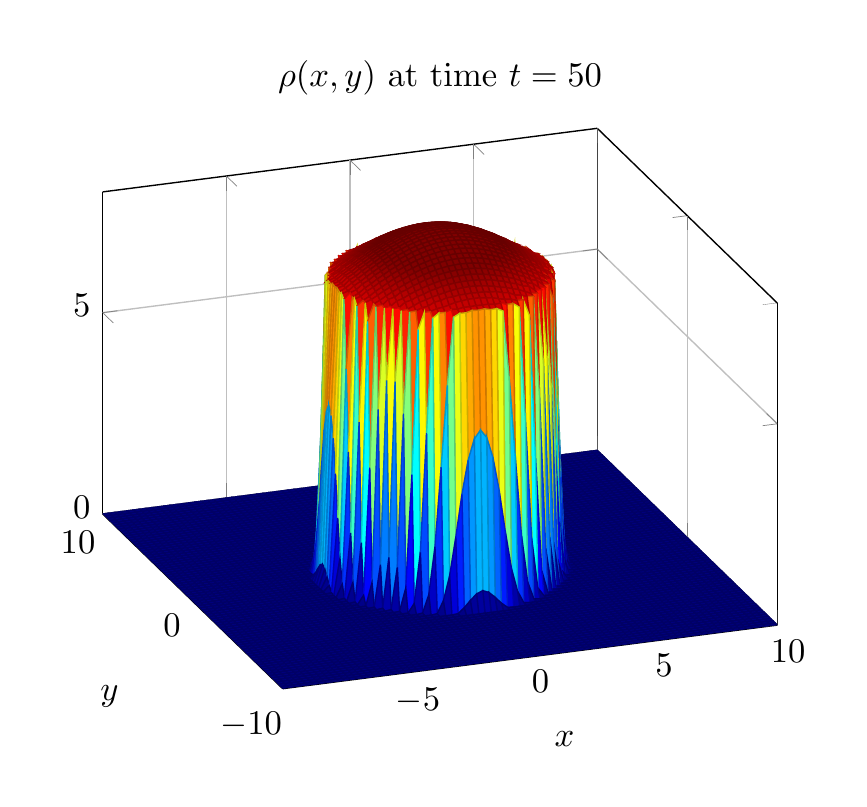}
  \includegraphics[scale=.65]{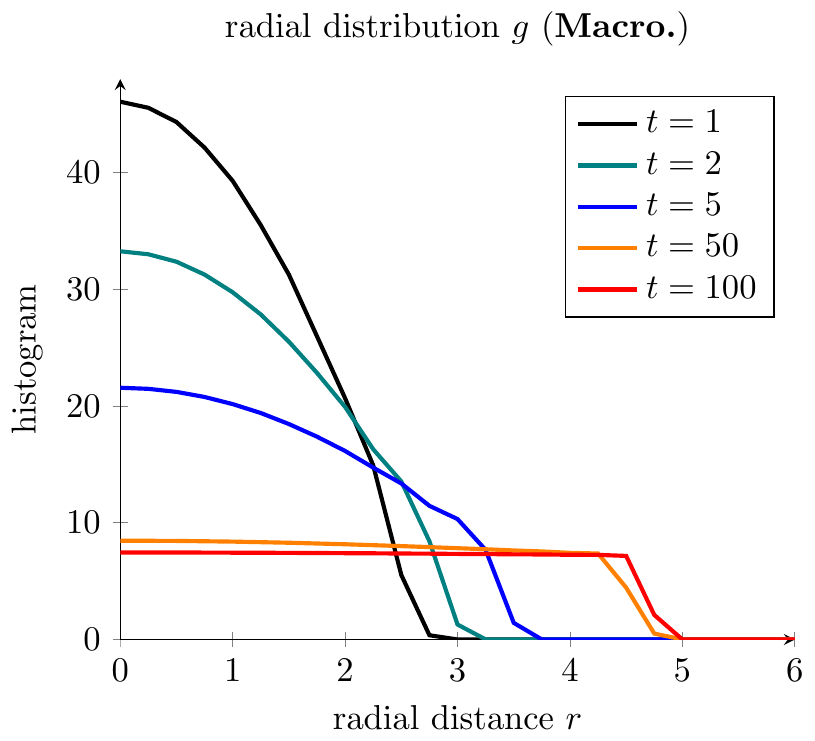}
  \caption{{\bf Left:} The solution $\rho({\bf x},t)$ of the macroscopic dynamics \eqref{eq:PDE_integral_sioux}\eqref{eq:G_rho_sioux} with initial condition \eqref{eq:rho_0} at $t=50$. In contrast to the {\it unstable} dynamics \eqref{eq:G_rho}\eqref{eq:PDE_integral}, the density $\rho$ is now converging to a stationary state with compact support. {\bf Right:} The radial distribution converges a plateau distribution. Parameters: $\Delta x=\Delta y=.25,\,\Delta t=5\cdot10^{-2}$. Total number of cells is $500$ with $R=.2$.}
  \label{fig:macroSioux_conv}
\end{figure}

Finally, we aim to compare the macroscopic dynamics with the microscopic dynamics with non-overlapping constraints (Fig.~\ref{fig:Micro_repulsion_growth_hulk}). With this aim, we need to find how to encompass non-overlapping constraints at the macroscopic scale which is the goal of the next section.

\begin{figure}[H]
  \centering
  \includegraphics[scale=.65]{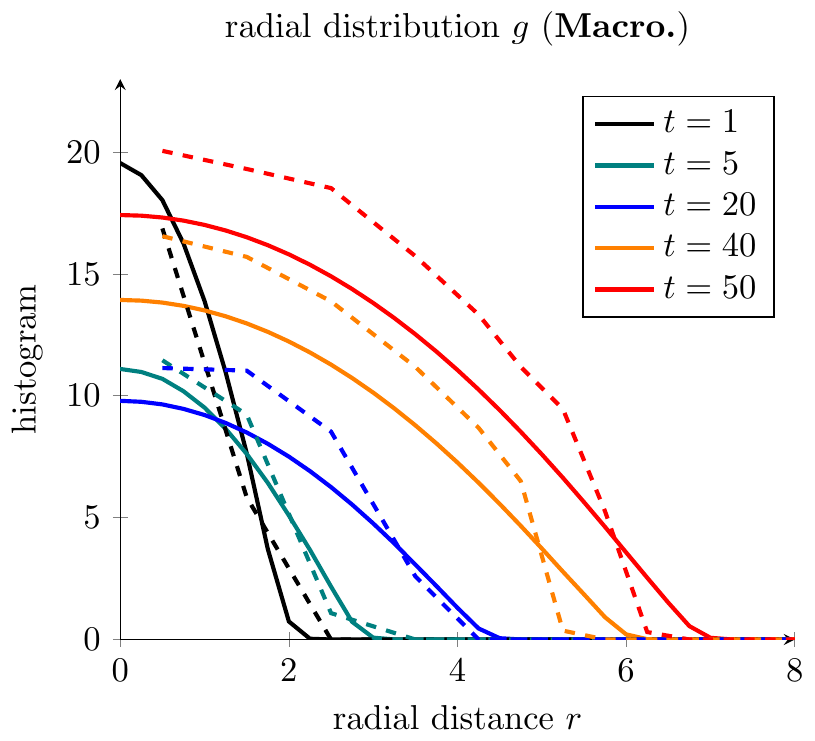}
  \includegraphics[scale=.65]{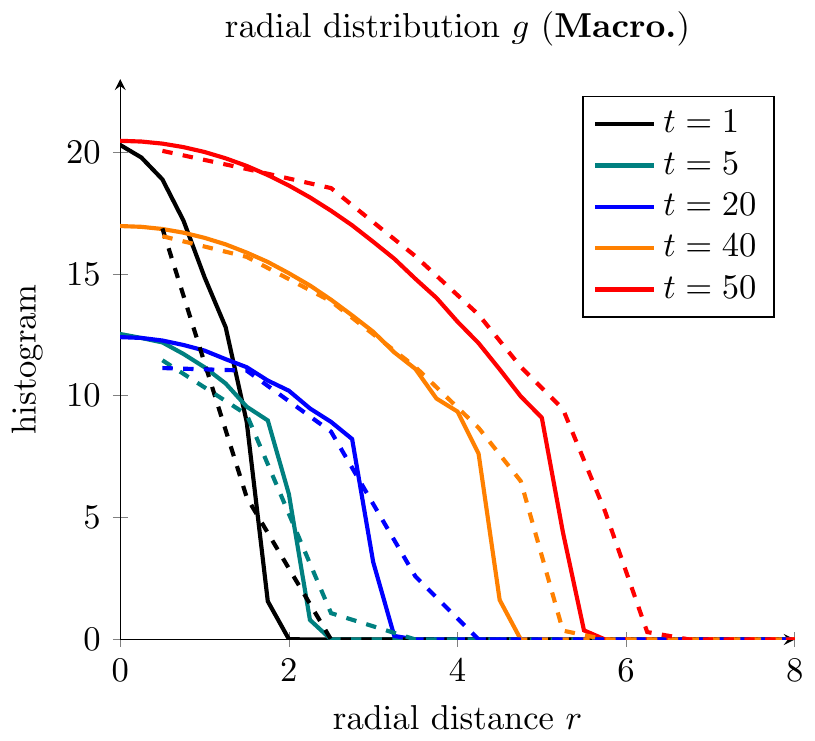}
  \caption{{\bf Left:} The solution $\rho({\bf x},t)$ of the macroscopic dynamics \eqref{eq:G_rho}\eqref{eq:PDE_integral} (plain) with growth rate $\mu=.05$ compare with the microscopic dynamics (dashed). {\bf Right:} Solution of the stabilized PDE \eqref{eq:PDE_integral_sioux}\eqref{eq:G_rho_sioux}. Parameters: $\Delta x=\Delta y=.25,\,\Delta t=5\cdot10^{-2}$. Total number of cells initially is $100$ with $R=.2$.}
  \label{fig:macroSioux_conv}
\end{figure}

\section{Asymptotic PDEs}

\subsection{Porous media equation ($R\to0$)}

As the cell radius $R$ is expected to be small compared to the characteristic length of the domain, we would like to investigate the asymptotic behavior of the macroscopic dynamics \eqref{eq:PDE_integral_sioux}\eqref{eq:G_rho_sioux} as $R$ tends to zero.

Using the change of variable ${\bf z}=({\bf x}-{\bf y})/2R$, we obtain:
\begin{displaymath}
  \overline{G}[\rho]({\bf x}) = (2R)^3\int_{{\bf z}\in \mathbb{R}^2} \phi(|{\bf z}|^2)\,{\bf z}\, {\bf h}\big(\rho({\bf x}-2R{\bf z})\big)\, \dd {\bf z}.
\end{displaymath}
Formally (${\bf h}$ is not a smooth function), we deduce:
\begin{eqnarray*}
  \overline{G}[\rho]({\bf x}) &=& (2R)^3\int_{{\bf z}\in \mathbb{R}^2} \phi(|{\bf z}|^2)\,{\bf z}\, \Big({\bf h}(\rho({\bf x})) -2R\,{\bf h}'(\rho({\bf x}))\nabla_{\bf x}\rho({\bf x}){\bf z}\Big) \big)\, \dd {\bf z} + \mathcal{O}(R^5) \\
                           &=& -(2R)^4\left(\int_{{\bf z}\in\mathbb{R}^2} \phi(|{\bf z}|^2){\bf z}\otimes {\bf z} \,\dd {\bf z}\right) {\bf h}'(\rho({\bf x})) \nabla_{\bf x} \rho({\bf x}) + \mathcal{O}(R^5),
\end{eqnarray*}
by symmetry of $\phi(|{\bf z}|^2)$. Polar coordinates yields:
\begin{displaymath}
  \int_{{\bf z}\in\mathbb{R}^2} \phi(|{\bf z}|^2){\bf z}\otimes {\bf z} \,\dd {\bf z} = \int_{r\geq0} \phi(|r|^2)r^3 \,\dd r \left[
    \begin{array}{ll}
      \pi & 0 \\
      0 & \pi
    \end{array}
  \right].
\end{displaymath}
Thus, we finally deduce:
\begin{displaymath}
  \overline{G}[\rho]({\bf x}) = \alpha_R {\bf h}'(\rho({\bf x}))\nabla_{\bf x} \rho({\bf x}) + \mathcal{O}(R^5),
\end{displaymath}
with $\alpha_R = \pi(2R)^4\int_{r\geq0} \phi(|r|^2)r^3\,\dd r$. Neglecting the high order terms in $R$, we deduce formally the following modified porous media equation:
\begin{equation}
  \label{eq:porous_modified}
  \partial_t \rho = \alpha_R\nabla_{\bf x}\cdot\big({\bf h}'(\rho)\rho\nabla_{\bf x} \rho\big) + \mu\rho.
\end{equation}
This equation does not have classical solution as ${\bf h}'$ is a discontinuous function at $\rho=\rho_*$. To avoid this discontinuity, one can 'smooth' the function ${\bf h}$ near $\rho_*$. Moreover, we can introduce the function $H$ satisfying:
\begin{equation}
  \label{eq:H}
  H'(\rho^2) = {\bf h}'(\rho).
\end{equation}
For the function ${\bf h}$ given by eq. \eqref{eq:h}, we obtain: $H(s)=(s-\rho_*^2)^+$. The Eq. \eqref{eq:porous_modified} becomes:
\begin{equation}
  \label{eq:porous_modified_2}
  \partial_t \rho = \frac{\alpha_R}{2}\Delta_{\bf x}H(\rho^2) + \mu\rho.
\end{equation}

{\bf remark}
  Without the constraint, the macroscopic PDE \eqref{eq:G_rho}\eqref{eq:PDE_integral} would lead to the classical porous media equation:
  \begin{displaymath}
    \partial_t \rho = \alpha_R\nabla_{\bf x}\cdot\big(\rho\nabla_{\bf x} \rho\big) + \mu\rho.
  \end{displaymath}



\subsection{Hele-Shaw equation}

\begin{figure}[H]
  \centering
  \includegraphics[scale=.5]{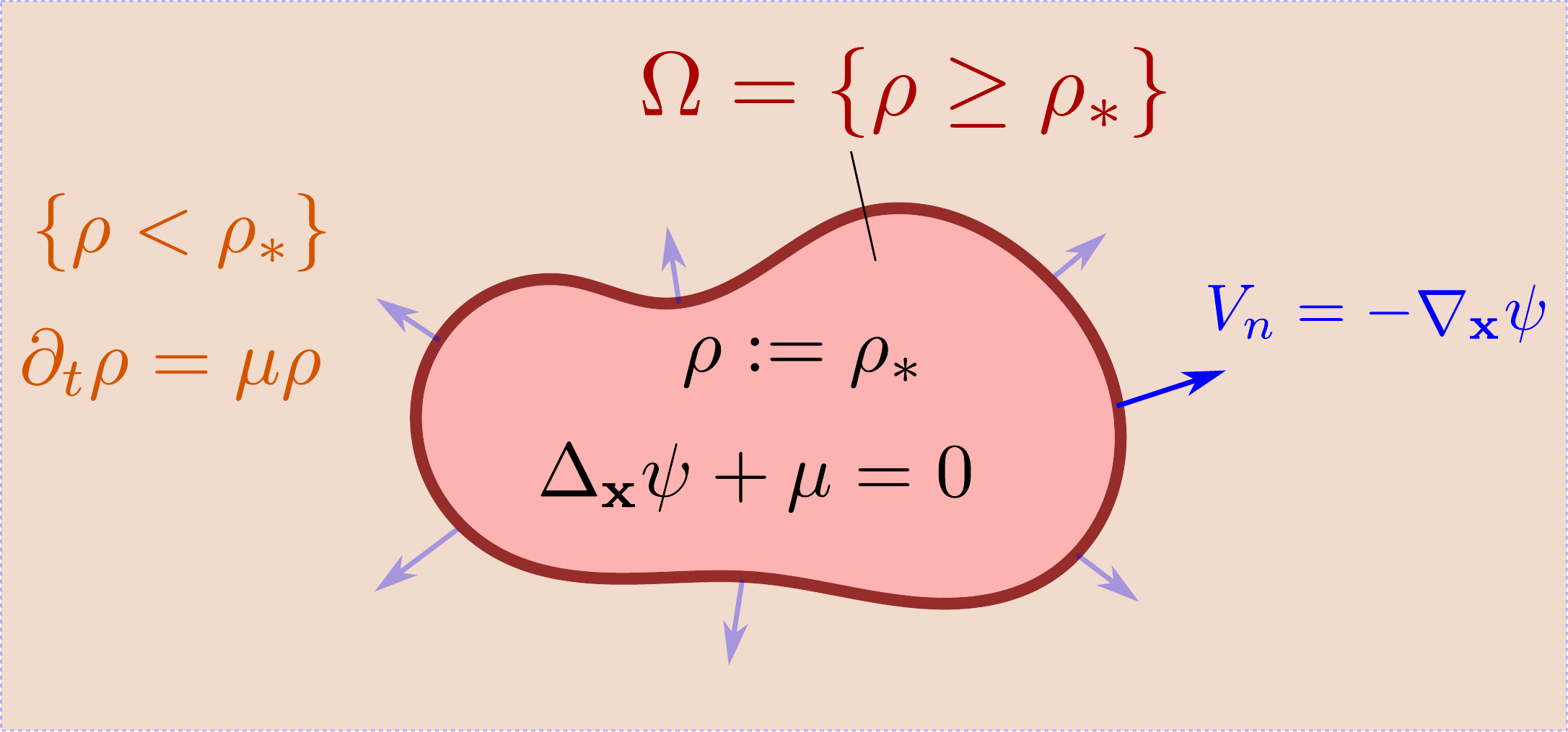}
  \caption{Illustration of the Hele-Shaw eq. \eqref{eq:rho_inf}-\eqref{eq:V}. The density $\rho$ is upper-bounded by the maximum packing threshold $\rho_*$. The region where the threshold is reached, denoted $\Omega$, is governed by the solution $\psi$ of an elliptic Eq. \eqref{eq:psi}.}
  \label{fig:schema_HS}
\end{figure}

\begin{figure}[H]
  \centering
  \includegraphics[scale=.6]{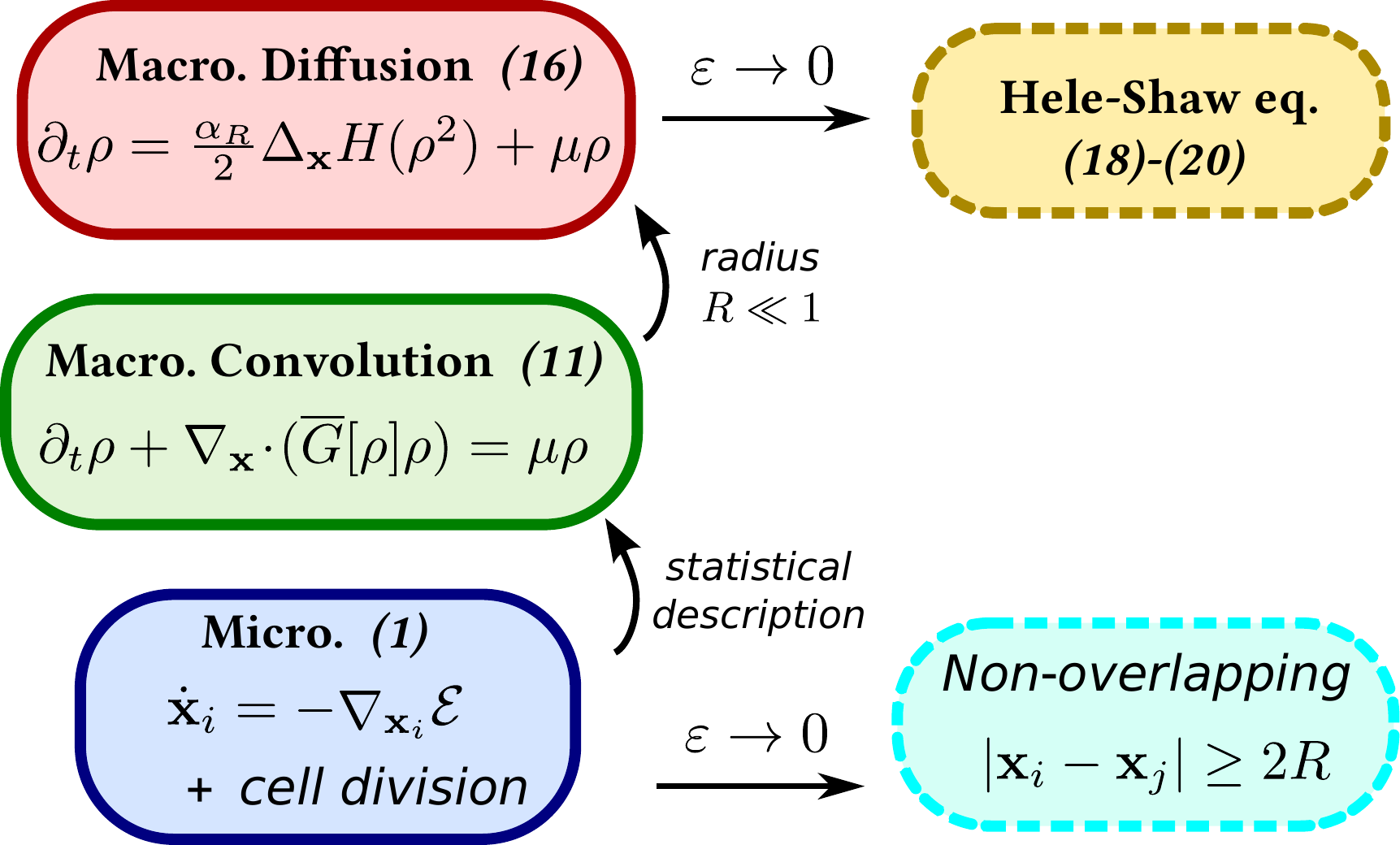}
  \caption{Schematic representation of the different models used to derive the Hele-Shaw equation. The starting point is the microscopic dynamics \eqref{eq:micro_kpp2}. The limit $\varepsilon\longrightarrow0$ denotes the asymptotic limit when the repulsion between cells becomes singular.}
  \label{fig:schema_3_models}
\end{figure}

We investigate the asymptotic limit of the dynamics when the repulsion between cells becomes 'infinite'. With this aim, we suppose that the repulsion function is of the form $\phi/\varepsilon$. The modified porous media Eq. \eqref{eq:porous_modified_2} becomes:
\begin{equation}
  \label{eq:porous_modified_limit}
  \partial_t \rho_\varepsilon = \frac{\alpha_R}{2\varepsilon}\Delta_{\bf x}H(\rho_\varepsilon^2) + \mu\rho_\varepsilon.
\end{equation}
We would like to identify the limit $\varepsilon\to0$ of the solution $\rho_\varepsilon$. Assuming that $\rho_\varepsilon \stackrel{\varepsilon \to 0}{\longrightarrow} \rho_\infty$, we can show formally that $\rho_\infty$ satisfies an Hele-Shaw type problem. We leave the proof in appendix \ref{sec:HS_proof}. The Hele-Shaw problem is as follow: consider the (moving) domain $\Omega(t)=\{{\bf x} \,|\,\rho({\bf x},t)\geq\rho_*\}$, then $\rho_\infty$ satisfies:
\begin{equation}
  \label{eq:rho_inf}
  \left\{
    \begin{array}{lc}
      \partial_t \rho_\infty = \mu \rho_\infty & \text{on } \mathbb{R}^2/\Omega, \\
      \rho_\infty = \rho_* & \text{on } \Omega.
    \end{array}
  \right.
\end{equation}
The evolution of the boundary of $\Omega$ is governed by a Laplace equation. Let $\psi$ be the solution of the following equation:
\begin{equation}
  \label{eq:psi}
  \left\{
    \begin{array}{cl}
       \Delta_{\bf x} \psi + \mu = 0 & \text{on } \Omega \\
      \psi =0 & \text{on } \partial\Omega
    \end{array}
  \right.
\end{equation}
Let ${\bf x}\in \partial\Omega$. The velocity of the (moving) boundary of $\Omega$ at ${\bf x} $ is given by (see Fig. \ref{fig:schema_HS}):
\begin{equation}
  \label{eq:V}
  V_n = -\frac{\rho_*}{\rho_*-\rho_0 \expo^{\mu t}} \nabla_{\bf x} \psi.
\end{equation}
In Fig. \ref{fig:schema_3_models}, we summarize the several intermediate models used to obtain the Hele-Shaw equation.

{\bf remark}
  As illustrated in Scheme \ref{fig:schema_3_models}, we need to consider two asymptotic limits in order to derive the Hele-Shaw equation: the radius of the cell should tend to zeros (i.e. $R\to0$) and the repulsion should become 'infinite' (i.e. $\varepsilon\to0$). Then, the Hele-Shaw equation is obtained as $\alpha_R/\varepsilon \to +\infty$. Since $\alpha_R = \mathcal{O}(R^4)$, we need to have $\varepsilon=o(R^4)$.

\subsection{Applications}

\subsubsection{Radial growth}

\begin{figure}[H]
  \centering
  \includegraphics[scale=.7]{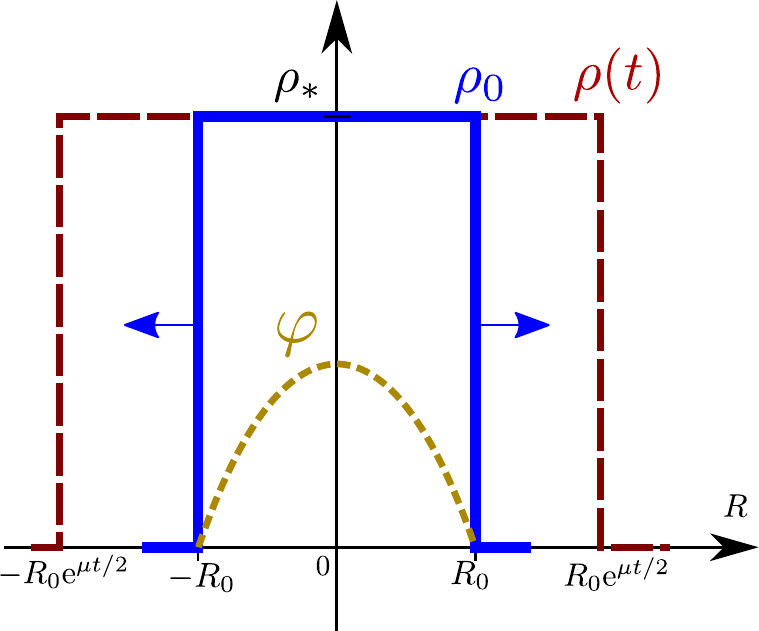}
  \includegraphics[scale=.7]{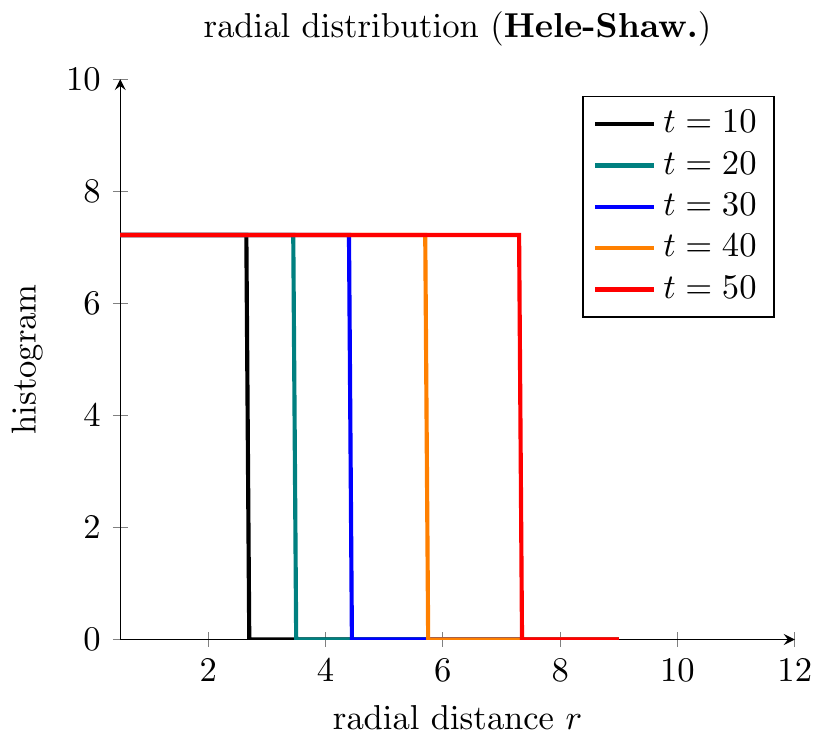}
  \caption{{\bf Left:} Schematic representation of the analytic solution of the Hele-Shaw eq. \eqref{eq:rho_inf}-\eqref{eq:V} when the initial condition is a unit disc. {\bf Right:} Evolution of the corresponding radial distribution to be compared with the microscopic dynamics (Fig.~\ref{fig:Micro_repulsion_growth_hulk}).}
  \label{fig:sol_1D}
\end{figure}

As a first test case for the Hele-Shaw Eq. \eqref{eq:rho_inf}-\eqref{eq:V}, we consider an initial condition where the density $\rho_0$ is uniformly distributed on a unit disc with density $\rho_*$. Thus, the region $\Omega$ is initially given by $\{x^2+y^2\leq1\}$. Then, we can solve explicitly the Hele-Shaw equation. Indeed, the solution of the elliptic Eq. \eqref{eq:psi} is given by:
\begin{displaymath}
  \psi(x,y) = \mu \frac{(x^2+y^2-R^2)}{4}.
\end{displaymath}
Thus, when $X$ is at the boundary of $\Omega$, we obtain the differential equation:
\begin{displaymath}
  X'=V_n(X) = -\nabla_{\bf x} \psi(X) = \frac{\mu}{2} X.
\end{displaymath}
The domain $\Omega_R$ will grow radially with: $R(t)=R_0 \expo^{\mu t/2}$. Notice that the total mass is growing exponentially with a factor $\mu$ as expected. We illustrate the solution on Fig. \ref{fig:sol_1D}. The evolution of the radial distribution of the density is similar to the one obtained for the microscopic dynamics (see fig.~\ref{fig:Micro_repulsion_growth_hulk}).

\subsubsection{Elliptic growth}

For our second illustration of the Hele-Shaw model, we use a more challenging initial condition with the density $\rho$ uniformly distributed on an ellipse::
\begin{displaymath}
  \Omega_0 = \left\{ \frac{x^2}{a^2} + \frac{y^2}{b^2} \leq 1\right\}.
\end{displaymath}
We can still solve explicitly the Hele-Shaw Eq. \eqref{eq:rho_inf}-\eqref{eq:V}. The solution of the elliptic Eq. \eqref{eq:psi} is given by:
\begin{displaymath}
  \psi(x,y) = \frac{\gamma \mu }{2} \left(\frac{x^2}{a^2} + \frac{y^2}{b^2} -1\right),
\end{displaymath}
with $\gamma= \frac{1}{1/a^2+1/b^2}$. Thus for $X=(x,y)\in\partial\Omega$:
\begin{equation}
  \label{eq:tp_x}
  X'=V(X) = -\nabla_{\bf x}  \psi(X) = \gamma\mu \left(
    \begin{array}{c}
      x/a \\
      y/b
    \end{array}
  \right).
\end{equation}
We can plug this expression to obtain the evolution of the contour $\partial\Omega$. Let $X(t)= (a(t)\cos \theta,\,b(t) \sin \theta)$ with $\theta$ fixed. Using eq. \eqref{eq:tp_x}, this ansatz leads to:
\begin{displaymath}
  \left(
    \begin{array}{c}
      a' \cos \theta \\
      b' \sin \theta
    \end{array}
\right) = \gamma\mu \left(
    \begin{array}{c}
      x/a^2 \\
      y/b^2
    \end{array}
\right) \quad \Rightarrow \quad \left\{
  \begin{array}{ccl}
    a' & = & \gamma\mu /a \\
    b' & = & \gamma\mu /b
  \end{array}
\right.
\end{displaymath}
Using the expression of $\gamma$, we obtain the following dynamical system for the minor/major axis of the ellipse:
\begin{equation}
  \label{eq:ODE_ab}
  \displaystyle a' = \mu  \frac{ab^2}{a^2+b^2} \quad,\quad
  b' = \mu \displaystyle \frac{a^2b}{a^2+b^2}
\end{equation}
Notice that the total mass of the solution $\rho_\infty$ growths exponentially with factor $\mu$ as expected. Indeed, let $m(t)=\int \rho_\infty\dd {\bf x} = \pi\cdot ab$. We have:
\begin{displaymath}
  \frac{d}{dt} m = \frac{d}{dt} \big(\pi ab\big) = \pi (a'b+ab') = \pi\mu \frac{ab^3+a^3b}{a^2+b^2} = \mu\pi\cdot ab = \mu\cdot m.
\end{displaymath}
{\bf remark}
  Another method to find explicitly the evolution of the ellipse consists in using elliptic coordinates $(s,\nu)$:
  \begin{displaymath}
    x= c \cosh s \cos \nu, \quad y= c \sinh s \sin \nu,
  \end{displaymath}
 where $c^2=a^2-b^2$ is the eccentricity of the ellipse. Suppose that only $s$ depends on time (e.g. $x(t)=c \cosh s(t) \cos \nu$), we find after simplification that:
  \begin{displaymath}
    s' = \frac{\mu}{2} \tanh s.
  \end{displaymath}
  Solving this differential equation gives the exact evolution of the domain $\Omega(t)$ in time.

\begin{figure}[H]
  \centering
  \includegraphics[scale=.8]{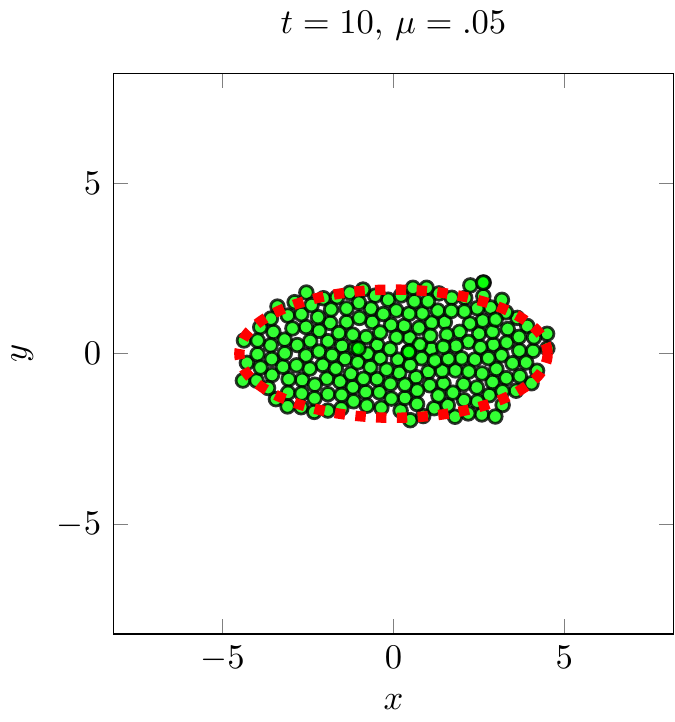} \quad
  \includegraphics[scale=.8]{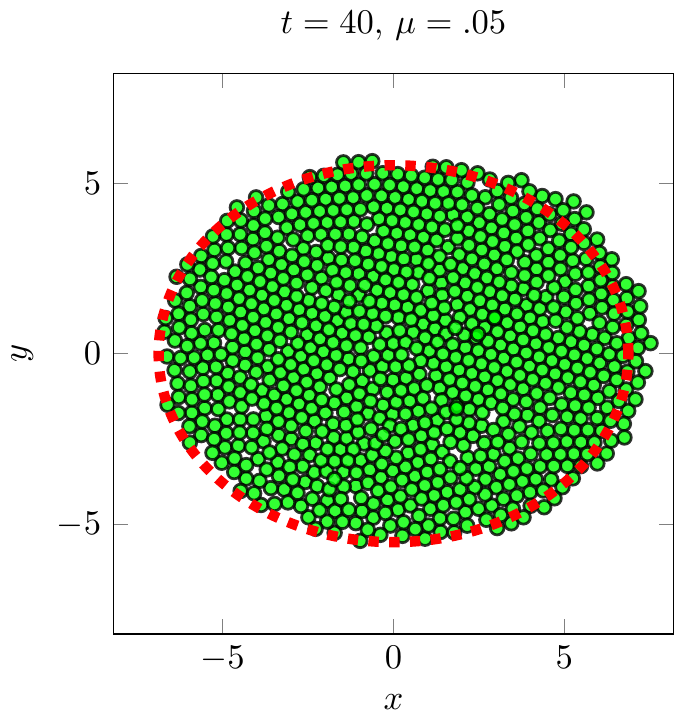}
  \caption{{\bf Left}: Initial condition for the second illustration of the Hele-Shaw problem. The (red) dashed line indicated the contour of the domain $\Omega$ obtained by solving the Hele-Shaw model \eqref{eq:rho_inf}-\eqref{eq:V}. {\bf Right}: Comparison of the solutions of the microscopic dynamics and Hele-Shaw problem at $t=40$.} 
  \label{fig:HS_ellipse}
\end{figure}

\begin{figure}[H]
  \centering
  \includegraphics[scale=.65]{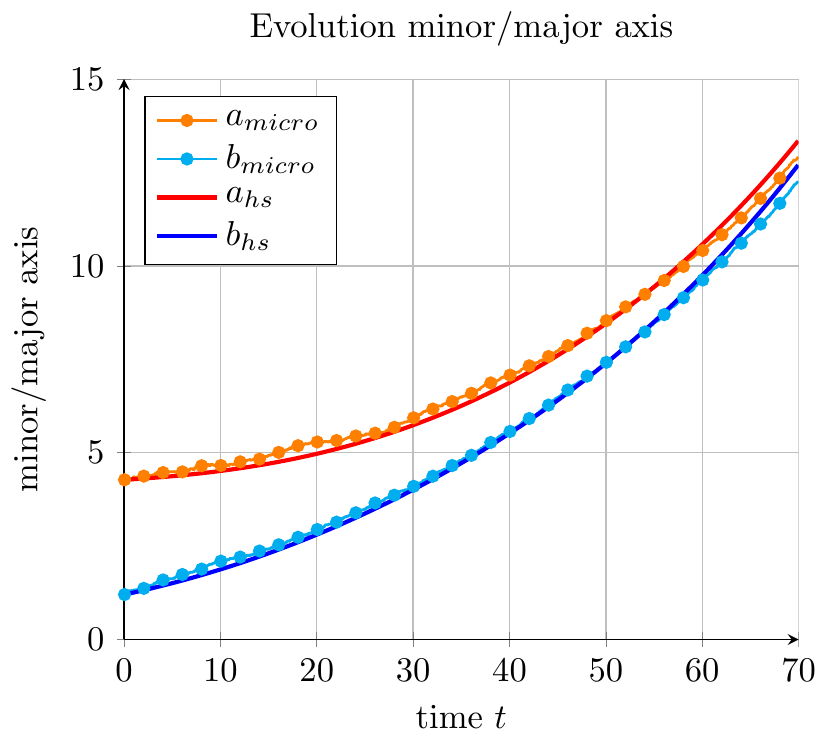} \quad
  \includegraphics[scale=.65]{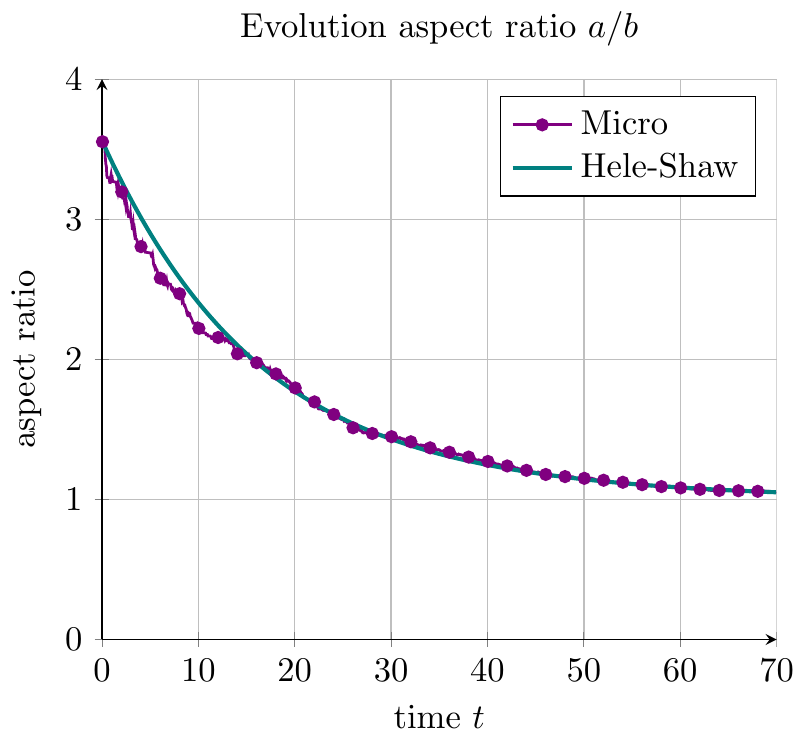}
  \caption{{\bf Left:} Evolution of the major/minor axis (resp. $a$ and $b$) starting from the ellipse for both microscopic dynamics \eqref{eq:micro_kpp2} and Hele-Shaw problem \eqref{eq:rho_inf}-\eqref{eq:V}. {\bf Right:} Evolution of the corresponding eccentricity: $a/b$.}
  \label{fig:ellipse_ab}
\end{figure}

To compare the solution of the Hele-Shaw problem with the microscopic dynamics \eqref{eq:micro_kpp2}, we perform a numerical simulation starting with $N=100$ cells distributed on an ellipse with major/minor axis $a=4$ and $b=1$ (see Fig. \ref{fig:HS_ellipse}). We let the microscopic dynamics \eqref{eq:micro_kpp2} evolve with the non-overlapping constraint (see section \ref{sec:micro_nonoverlapping}). At each time step, we estimate the major/minor axis denoted $a(t)$, $b(t)$. The estimation is performed by doing a Principal Component Analysis on the cloud of points $\{x_i(t),y_i(t)\}_i$. 
We compare the evolution of $a(t)$ and $b(t)$ with the one predicted by the solution of the Hele-Shaw model (Fig. \ref{fig:ellipse_ab}-left). We observe qualitatively the same behavior: both curves increase exponentially and $b$ is getting closer to $a$. The later observation means that the shape of the tumor is getting closer to a disc as we observe in Fig. \ref{fig:HS_ellipse}.

After time $t=60$ time units, we start to observe some discrepancies, the solution of the Hele-Shaw increasing faster than the one of the microscopic model. One could explain this difference by the numerical errors of the microscopic dynamics. The constraint of non-overlapping (i.e. $|{\bf x}_j-{\bf x}_i|\geq2R$) are not perfectly satisfied and thus the microscopic distribution is (slightly) less spread than what it should. These small errors add up as the number of cells grow, leading to the qualitative difference between the microscopic and macroscopic models. Nonetheless, if we only look at the aspect ratio of the ellipse (i.e. $a/b$), we observe an almost perfect agreement between the solution of the Hele-Shaw equation and the one of the microscopic model (Fig. \ref{fig:ellipse_ab}-right). This is quite remarkable as several simplifications have been done (e.g. $R\ll 1$, $\varepsilon\to0$) to derive the Hele-Shaw equation.

\section{Conclusion}


In this work, we have studied an agent-based model for tumor growth, referred to as the {\it microscopic} dynamics, modeling cell-cell repulsion and cell division. The associated macroscopic equation did not capture the large time behavior of the microscopic dynamics. For instance, when cell division is turned off, the solutions of the microscopic model converge to a stationary state compactly supported, whereas the solutions of the macroscopic model keep spreading in space. The discrepancy between the two models can be explained by the instability of Dirac masses for the macroscopic equation. To obtain a macroscopic model in accordance with the microscopic dynamics, we proposed a modified version of the macroscopic equation introducing a congestion feature in the interaction kernel. We showed that this model led to a modified porous media equation where the diffusion is only active in regions of large density. We confirmed the relevance of the proposed model by comparing numerically the simulations of both micro- and macro- dynamics. We also investigated the asymptotic limit of the dynamics when the repulsion between cells becomes singular (leading to non-overlapping constraints in the microscopic model). We showed (formally) that in this limit, we obtain a Hele-Shaw type problem at the macroscopic level. We numerically confirmed the relevance of this macroscopic dynamics by comparing solutions of the microscopic models with explicit solutions of the Hele-Shaw problem.

On the mathematical viewpoint, this work is a first attempt to validate macroscopic models for tumor growth with a microscopic dynamics. This work shows that Hele-Shaw type problems -first derived in \cite{mellet_hele-shaw_2015} for tumor growth models- can be obtained in a well-chosen asymptotics of a microscopic dynamics. These results show the relevance of the use of such models for the study of biological tumors.

On the biological viewpoint, this work raises exciting perspectives for the study of the mechanisms of glioblastoma. Future works will aim at studying the influence of a vascular network on tumor expansion, at the microscopic and macroscopic scales. These models will be validated through comparison with experimental data. The hope is to be able to identify the mechanisms of glioma cell invasion in brain tumor, and detect different invasion patterns. 

Further perspectives include rigorously proving the derivation of the Hele-Shaw problem from the microscopic model. The difficulty is that many asymptotic limits are necessary (e.g. large number of cell $N$, small radius $R$, singular limit for the kernel $\phi$). One could also extend the model to describe more complex cell division dynamics with for example a growth rate that depends on the density. This could lead to a Hele-Shaw type problem with non-linear elliptic equation.

\bibliographystyle{plain}
\bibliography{biblio_repulsion}

\clearpage

\appendix

\section{Numerical schemes}
\label{sec:appendix_A}

\subsection{Particle dynamics}

To discretize the microscopic dynamics \eqref{eq:micro_kpp2}, we use an Euler method with an adaptive time step $\Delta t$ to ensure that the energy \eqref{eq:energy_micro} is decaying:
\begin{itemize}
\item[1)] Let $\{x_i(t_n)\}_i$, $\mathcal{E}(t_n)$ the corresponding energy and a time step $\Delta t$
\item[2)] For all $i\in\{1,N\}$, compute:
  \begin{displaymath}
    x_i(t_{n+1}) = x_i(t_{n}) - \Delta t\!\!\sum_{j=1,j\neq i}^N \phi_{ij}\cdot\big(x_j(t_n)-x_i(t_n)\big)
  \end{displaymath}
  Deduce the corresponding energy $\mathcal{E}(t_{n+1})$.
\item[3)] If $\mathcal{E}(t_{n+1})>\mathcal{E}(t_{n})$, go back to 2) with $\widetilde{\Delta t}=\Delta t/2$.\\
  Otherwise update $t_n$ to $t_{n+1}$.
\end{itemize}

\subsection{PDE dynamics}

We use an upwind-method to solve Eq. \eqref{eq:G_rho}-\eqref{eq:PDE_integral}. To simplify the notation, we illustrate the method in the one dimensional case. We use a uniform grid in space and time and denote: $\rho_i^n= \rho(x_i,t^n)$ with $x_i=i\Delta x$ and $t^n=n\Delta t$. The scheme is based on the following discretization:
\begin{displaymath}
  \frac{\rho_i^{n+1} - \rho_i^n}{\Delta t} + \frac{(\rho G)_{i+1/2}-(\rho G)_{i-1/2}}{\Delta x} = 0,
\end{displaymath}
where $(\rho G)_{i+1/2}$ is the value of $\rho(x)G[\rho](x)$ at the interface $x_{i+1/2}$. To estimate this value, we first estimate the velocity at the interface $x_{i+1/2}$: $G_{i+1/2}= \frac{G_{i}+G_{i+1}}{2}$. Then, we decentralize:
\begin{displaymath}
  (\rho G)_{i+1/2} = \left\{
    \begin{array}{ll}
      \rho_i G_{i+1/2} & \text{if } G_{i+1/2} >0, \\
      \rho_{i+1} G_{i+1/2} & \text{if } G_{i+1/2} <0. \\
    \end{array}
  \right.
\end{displaymath}
The CFL condition associated with this scheme is given by $\lambda\Delta t/\Delta x$ where $\lambda=2\max |G|$.

For the porous media Eq. \eqref{eq:porous_modified_2}, we use the formulation (in 1D) $\partial_t\rho = \alpha_R \partial_{xx} (\rho^2)/2$ to deduce
\begin{displaymath}
  \frac{\rho_i^{n+1} - \rho_i^n}{\Delta t} = \frac{\alpha_R}{2} \frac{ (\rho_{i+1}^{n})^2 - 2(\rho_i^{n})^2 + (\rho_{i-1}^{n})^2}{\Delta x^2}.
\end{displaymath}

In our all simulations, we have verified that both positivity and energy decaying were satisfied. A more sophisticated scheme has been proposed in \cite{carrillo_finite-volume_2015} that can guarantee both properties.

\section{Derivation Hele-Shaw equations}
\label{sec:HS_proof}

We derive formally the Hele-Shaw starting from the system:
\begin{equation}
  \label{eq:porous_modified_limit_bis}
  \partial_t \rho_\varepsilon = \frac{\alpha_R}{2\varepsilon}\Delta_{\bf x}H(\rho_\varepsilon^2) + \mu\rho_\varepsilon.
\end{equation}
The goal is to find the asymptotic limit as $\varepsilon\to0$. With this aim, we assume that a limit $\rho_\varepsilon \stackrel{\varepsilon \to 0}{\longrightarrow} \rho_\infty$ exists and we try to identify what equation $\rho_0$ satisfies. 

Multiplying Eq. \eqref{eq:porous_modified_limit_bis} by $\varepsilon$ and passing to the limit $\varepsilon\to0$, we deduce:
\begin{equation}
  \label{eq:limit_D_rho}
  \Delta_{\bf x} H(\rho_\infty^2) = 0.
\end{equation}
Thus, using the notation $\Omega=\{\rho_\infty\geq\rho_*\}$, we deduce that $\Delta\rho_\infty^2=0$ on $\Omega$ and therefore $\rho_\infty$ is constant on $\Omega$. Assuming that $\rho_\infty$ is {\it continuous}, we deduce:
\begin{equation}
  \label{eq:rho_1}
  \rho_\infty=\rho_* \qquad \text{on } \Omega.
\end{equation}

Now for any ${\bf x}$ in $\mathbb{R}^n\backslash \Omega$ (i.e. $\rho_\infty({\bf x})<\rho_*$), we suppose point-wise convergence of $\rho_\varepsilon$ to $\rho_\infty$ thus for $\varepsilon$ small enough $\rho_\varepsilon({\bf x})<\rho_*$ and therefore $H(\rho_\varepsilon^2)=0$. Therefore \eqref{eq:porous_modified_limit_bis} reduces to:
\begin{displaymath}
  \partial_t \rho_\varepsilon =   \mu\rho_\varepsilon \qquad \text{on } \mathbb{R}^n\backslash \Omega,
\end{displaymath}
for $\varepsilon$ small enough. Passing to the limit, we obtain: $\partial_t \rho_\infty =   \mu\rho_\infty$ on $\mathbb{R}^n\backslash \Omega$ and thus:
\begin{equation}
  \label{eq:rho_exp}
  \rho_\infty=\expo^{\mu t}\rho_0 \qquad \text{on } \mathbb{R}^n\backslash \Omega.
\end{equation}
Combining \eqref{eq:rho_1} and \eqref{eq:rho_exp} give:
\begin{equation}
  \label{eq:struc_rho}
  \rho_\infty({\bf x}) = \rho_* \chi_{\Omega}({\bf x}) + (1-\chi_\Omega({\bf x})) \expo^{\mu t}\rho_0({\bf x}),
\end{equation}
where $\chi_\Omega$ is the indicator function of the set $\Omega$.

To identify how the frontier of the set $\Omega$ moves, we perform a perturbation analysis near $\rho_\infty$. We suppose the following ansatz:
\begin{equation}
  \label{eq:Hilbert}
  \rho_\varepsilon = \rho_\infty + \varepsilon \rho_1 + \mathcal{O}(\varepsilon^2).
\end{equation}
Notice that on $\mathbb{R}^n\backslash \Omega$, $\rho_\varepsilon=\rho_\infty$ and therefore:
\begin{equation}
  \label{eq:rho_1_outside}
  \rho_1=0 \qquad \text{on } \mathbb{R}^n\backslash \Omega.
\end{equation}
Plug in the ansatz into the Eq. \eqref{eq:porous_modified_limit_bis}:
\begin{eqnarray*}
  \label{eq:plugin}
  \partial_t \rho_\varepsilon &=& \frac{\alpha_R}{2\varepsilon} \Delta_{\bf x} H\big((\rho_\infty+\varepsilon\rho_1+...)^2\big) + \mu\rho_\varepsilon \\
          &=& \frac{\alpha_R}{2\varepsilon} \Delta_{\bf x} \Big(H(\rho_\infty^2) + 2\rho_\infty\varepsilon H'(\rho_\infty^2)\rho_1\big) + \mu\rho_\varepsilon + \mathcal{O}(\varepsilon).
\end{eqnarray*}
Using that $H'(\rho_\infty^2)=\chi_\Omega$ and that the support of $\rho_1$ is on $\Omega$, we deduce $H'(\rho_\infty^2)\rho_1=\rho_1$. Thus,
\begin{displaymath}
  \partial_t \rho_\varepsilon = \alpha_R \rho_\infty \Delta_{\bf x} \rho_1 + \mu\rho_\varepsilon + \mathcal{O}(\varepsilon).
\end{displaymath}
Passing to the limit $\varepsilon\to0$ gives:
\begin{equation}
  \label{eq:limit_rho_1}
  \partial_t \rho_\infty = \alpha_R \rho_\infty \Delta_{\bf x} \rho_1 + \mu\rho_\infty.  
\end{equation}

It remains to identify the perturbation $\rho_1$ inside the domain $\Omega$. Since $\rho_\infty=\rho_*$ on $\Omega$, Eq. \eqref{eq:limit_rho_1} gives:
\begin{equation}
  \label{eq:psi_1}
  0 = \alpha_R \rho_* \Delta_{\bf x} \rho_1 + \mu\cdot\rho_* \qquad \text{on } \Omega.
\end{equation}
Thus, denoting $\psi$ the solution to the elliptic equation:
\begin{equation}
  \label{eq:elliptic_rho1}
  \left\{
    \begin{array}{cccl}
      \Delta_{\bf x} \psi + \mu &=& 0 & \qquad \text{in } \Omega \\
      \psi &=& 0 & \qquad \text{in } \partial\Omega. 
    \end{array}
  \right.
\end{equation}
we have $\alpha_R \rho_1 = \psi$. We now combine the computations above to conclude. Taking the time derivative of \eqref{eq:struc_rho}, we obtain:
\begin{eqnarray*}
  \partial_t \rho_\infty &=& (\rho_*-\expo^{\mu t}\rho_0)\partial_t \chi_\Omega + (1-\chi_\Omega)\mu \expo^{\mu t}\rho_0\\
          &=& (\rho_*-\expo^{\mu t}\rho_0)\partial_t \chi_\Omega + (1-\chi_\Omega)\mu \rho_\infty,
\end{eqnarray*}
since $\rho_\infty=\expo^{\mu t}\rho_0$ on $\mathbb{R}^n\backslash \Omega$. Using eq. \eqref{eq:limit_rho_1}, we deduce:
\begin{displaymath}
  (\rho_*-\expo^{\mu t}\rho_0)\partial_t \chi_\Omega + (1-\chi_\Omega)\mu \rho_\infty = \rho_\infty \Delta_{\bf x} \psi + \mu\rho_\infty,
\end{displaymath}
leading to:
\begin{equation}
  \label{eq:almost_there}
  (\rho_*-\expo^{\mu t}\rho_0)\partial_t \chi_\Omega = \rho_* (\Delta_{\bf x} \psi + \mu \chi_\Omega),
\end{equation}
using \eqref{eq:rho_1}. Formally, $\partial_t \chi_\Omega= V_n\cdot \mathcal{H}^{1}|_{\partial\Omega}$ where $V_n$ is the normal velocity of $\partial\Omega$ and $\mathcal{H}^{1}$ the one-dimensional Hausdorff measure. Moreover (see \cite{mellet_hele-shaw_2015}), we also have that $\Delta_{\bf x} \psi + \mu \chi_\Omega = |\nabla_{\bf x} \psi|\cdot\mathcal{H}^{1}|_{\partial\Omega}$. Therefore, we deduce:
\begin{equation}
  \label{eq:V_n}
  V_n = -\frac{\rho_*}{\rho_*-\expo^{\mu t}\rho_0} \nabla_{\bf x}\psi,
\end{equation}
which corresponds to a Hele-Shaw free boundary problem.

{\bf remark}
  To prove rigorously the limit, we need to show that $\psi\geq0$ on the domain $\Omega$. We would deduce that: $H(\rho_\varepsilon) = \rho_\varepsilon$ on $\Omega$.

\end{document}